\documentclass[10pt]{article}

\usepackage{amsmath}
\usepackage{subfigure}
\usepackage{amssymb}

\usepackage{subfig} 
\usepackage{epsfig}
\usepackage{caption}

\usepackage{graphicx}

\usepackage{cite}

\usepackage{color}

\usepackage{a4wide}

\usepackage[labelfont=bf,labelsep=period,justification=raggedright]{caption}

\makeatletter
\renewcommand{\@biblabel}[1]{\quad#1.}
\makeatother

\date{}

\newcommand{\dd}[0]{{\textrm d}}
\newcommand{\Ca}[0]{{\textrm{[Ca}^{2+}]_i}}

\newcommand{\etal}[0]{{\emph{et al.}}}

\begin{document}
\begin{flushleft}

{\Large \textbf{The role of cellular coupling in the spontaneous generation of electrical activity in uterine tissue}}

\vspace{0.4cm}
{Jinshan Xu$^{1}$,
Shakti N. Menon$^{2}$,
Rajeev Singh$^{2}$,
Nicolas B. Garnier$^3$,
Sitabhra Sinha$^{2}$ and
Alain~Pumir$^{3,4,\ast}$}

\vspace{0.2cm}
\bf{1} College of Computer Science, Zhejiang University of Technology, Hangzhou, China. \\
\bf{2} The Institute of Mathematical Sciences, C.~I.~T. Campus, Taramani, Chennai~600113, India. \\
\bf{3} Laboratoire de Physique, Ecole Normale Sup\'erieure de Lyon, F-69007, Lyon, France. \\
\bf{4} Max-Planck Institute for Dynamics and Self-Organisation, D-37073, G\"ottingen,~Germany.

\vspace{0.1cm}
$\ast$ Corresponding author: alain.pumir@ens-lyon.fr\\
Telephone: +33-472-728-138, Fax: +33-472-728-080
\end{flushleft}

\thispagestyle{empty}

\section*{Abstract}

The spontaneous emergence of contraction-inducing electrical activity in the uterus at the beginning of labor remains poorly understood, partly due to the seemingly contradictory observation that isolated uterine cells are not spontaneously active. It is known, however, that the expression of gap junctions increases dramatically in the approach to parturition, which results in a significant increase in inter-cellular electrical coupling. In this paper, we build upon previous studies of the activity of electrically excitable smooth muscle cells (myocytes)
and investigate the mechanism through which the coupling of these cells to 
electrically passive cells results in
the generation of spontaneous activity in the uterus. Using a recently 
developed, realistic model of uterine muscle cell dynamics, we investigate a system consisting of a myocyte coupled to passive cells. We then extend our analysis to a simple two-dimensional lattice model of the tissue, with each myocyte being coupled to its neighbors, as well as to a random number of passive cells.  We observe that different dynamical regimes can be observed over a range of gap junction conductances: at low coupling strength, the activity is confined to cell clusters, while the activity for high coupling may spread across the entire tissue. Additionally, we find that the system supports the spontaneous generation of spiral wave activity. Our results are both qualitatively and quantitatively consistent with observations from \textit{in vitro} experiments. In particular, we demonstrate that an increase in inter-cellular electrical coupling, for realistic parameter values, strongly facilitates the appearance of spontaneous action potentials that may eventually lead to parturition.

\section*{Author Summary}

Preterm births, induced by premature contractions, may lead to irreversible infant neurological damage, and also account for a significant number of perinatal deaths. A major impediment in the treatment of this major public health issue is the fact that the mechanism by which uterine contractions commence is poorly understood. One particularly puzzling aspect of this process is that although the organ spontaneously develops contraction-inducing electrical activity, none of its constituent cells are independently capable of doing so. In other words, the uterus does not have any pacemaker, contrary to the situation observed in the heart, for instance. What, then, causes spontaneous uterine contractions? A clue may lie in the well-documented fact that gap junction expression increases very significantly shortly before term, resulting in stronger inter-cellular coupling. We show, using a recently developed, realistic computational model, with parameter values consistent with reported data, that this increased coupling can lead to spontaneous activity, both at the level of an isolated muscle cell and in a simple lattice model of uterine tissue. The elucidation of this process may provide an important insight into the genesis of preterm contractions, and may suggest possible pathways into the treatment of this widespread pathology.


\section{Introduction}
\label{sec:intro}

It is well known that the contraction of uterine smooth muscle cells
(myocytes) is triggered by electrical activity resulting from action
potentials that evolve from single spikes to spike trains in the lead
up to parturition~\cite{Landa1959,Csapo1963,Bengtsson1984}. However,
the precise mechanism underlying the transition of the uterus from the
quiescent organ seen during most stages of pregnancy, to the
rhythmically contracting muscle observed at the onset of labor,
remains to be fully explained. Preterm births, which occur prior to 37
weeks of gestation, can spontaneously arise from early, undesired
uterine contractions~\cite{Pervolaraki2012}, and hence this process is
highly significant from a clinical perspective. Indeed, recent data
suggests that preterm births constitute approximately 10\% of all
births~\cite{Martin2009}, and that rates of spontaneous preterm labor
have been increasing at the same rate as elective or induced preterm
births~\cite{Norman2009}. Preterm births have been implicated as the
cause of over a million neonatal deaths per year worldwide, and in
around 50\% of all cases of infant neurological
damage~\cite{Pervolaraki2012}. In the USA alone, they have been linked
to 40\% of all infant deaths~\cite{Mathews2012}. A clearer
understanding of the mechanism of spontaneous uterine tissue
contraction could therefore greatly facilitate the development of
effective strategies to help curb neonatal mortality and morbidity.

It has been postulated that spontaneous electrical oscillations in
uterine tissue, observed prior to the mechanical contractions of the
pregnant uterus, may be initiated by ``pacemaker''
cells~\cite{Garfield2007,Lammers2013}, similar to the Interstitial
cells of Cajal, which are known to act as a pacemaker in other smooth
muscles, such as the rabbit urethra~\cite{Sergeant2000,McHale2006}.
However, despite much effort aimed at identifying the origin of
spontaneous uterine contractions~\cite{Wray2001,Duquette2005}, there
has thus far been no clear evidence for the existence of such cells in
the uterus. On the contrary, uterine tissue is known to contain 
an
abundance of electrically passive cells, such as
Interstitial Cajal-like Cells (ICLCs) or telocytes~\cite{Popescu2010}
which, despite their similarity to Cajal cells,
have been argued to inhibit electrical
activity~\cite{McHale2006,Popescu2007}. 
ICLCs have a density of about $100-150$~cells/$\rm{mm}^2$ in the
uterus, and contribute up to $\sim$18\% of the cell population
immediately below the mucosal
epithelium~\cite{Popescu2005,Popescu2007}. Their density is
highest on the surface of the uterus and decreases to
a value of around $7.5\%$  in
muscularis~\cite{Popescu2007}. Other electrically passive uterine cells
include fibroblasts~\cite{Duquette2005,Popescu2007}, which play
an important role in remodeling the human uterine cervix during
pregnancy and parturition~\cite{Takemura2005, Malmstrom2007} despite
their small population~\cite{Duquette2005}.

The hypothesis that spontaneous electrical behavior is an
inherent property of uterine smooth-muscle cells~\cite{Garfield2007}
has gained traction as it is known that numerous electrophysiological
changes occur in the myometrium during the course of pregnancy. In
particular, the outward (K$^{+}$) and inward (Na$^{+}$, as well as
Ca$^{2+}$) currents are known to
change~\cite{Yoshino1997,Wang1998}. 
It has been observed from
experiments on rat uterine myocytes that the recorded peak current in
the Na$^{+}$ channel increases from $\sim2.8 \mu \rm{A}/\rm{cm}^2$ in
a non-pregnant uterus to $\sim5.1 \mu \rm{A}/\rm{cm}^2$ at late
pregnancy, while the corresponding peak current in the Ca$^{2+}$
channel decreases from $\sim5.7 \mu \rm{A}/\rm{cm}^2$ to $\sim3.4 \mu
\rm{A}/\rm{cm}^2$ over the same range~\cite{Yoshino1997}.
Moreover, the myocyte resting potential has been observed to
change from a value close to $-70 mV$ at the beginning of pregnancy
to around $-55 mV$ at midterm~\cite{Parkington1999}. These
changes can be related to the morphological modulations of the uterine
tissue~\cite{Bengtsson1984}, that are particularly apparent
shortly before delivery. As the tissue enlarges to accommodate the
growing fetus, its weight increases from around $75$g to $1300$g in
humans~\cite{Young2007}. This is accompanied by changes in both the
surface area of a single myocyte cell, from $\sim1930\mu
\rm{m}^2$ to $\sim7600 \mu \rm{m}^2$ during late pregnancy, and,
consistent with the observed hypertrophy of the uterus, a five-fold
increase in the cell capacitance~\cite{Yoshino1997}. However,
it has not yet been demonstrated that such changes are
responsible for the spontaneous generation of action potentials
necessary for the periodic mechanical contractions of the uterine
tissue~\cite{Young2007}.

An alternative paradigm for the genesis of coherent uterine
activity is hinted at by the
fact that an even more dramatic change occurs close to term in
the uterus. The fractional area of gap junctions, defined as the
ratio of the membrane area occupied by gap junctions to the total
membrane area, has been observed to show a $20$-fold increase in the
rat uterus~\cite{Miller1989}. Furthermore, the gap junctional
conductance has been found to increase from $\sim4.7$ nS at normal
preterm to $\sim32$ nS during delivery~\cite{Miyoshi1996}, while a
reduced expression of the major gap junction protein connexin 43 in
transgenic mice is known to significantly delay
parturition~\cite{Doring2006}. The importance of gap junction
expression is manifested most spectacularly in the observation that
chemical disruption of the gap junctions immediately inhibits the
oscillatory uterine
contractions~\cite{Tsai1998,Wang2002,Loch-Caruso2003}. These findings
strongly suggest that gap junctional coupling between proximate cells
plays a very important role in the development of coordinated
uterine electrophysiological activity, and may be responsible for the
transition from the weak, desynchronized myometrial contractions seen
in a quiescent uterus to the strong, synchronous contractions observed
during labor~\cite{Miller1989,Miyoshi1996}.

It has recently been observed that the coupling of an excitable cell to an electrically passive cell in a simple theoretical model of myocyte activity can give rise to oscillations, even if neither of the cells are initially oscillating~\cite{Jacquemet2006}. This prediction seems to be borne out by experiments, as the coupling of electrically active and passive cells in an assembly is indeed known to significantly affect the observed synchronization 
dynamics~\cite{Kryukov2008,Majumder2012}, while complicated dynamical regimes are observed in preparations of weakly coupled cardiac myocytes~\cite{Bub2002,Pumir2005}. The physiological significance of this phenomenon can be inferred from the fact that the activity synchronizes at high coupling strengths in both experimental preparations and numerical simulations of theoretical models. This synchronization first occurs over small regions (cell clusters) whose size gradually increases to fill out the full media, so that all cells beat with the same frequency~\cite{Kryukov2008,Chen2009,Singh2012}.

Consequently, it has been hypothesized~\cite{Singh2012} that
spontaneous oscillatory behavior could be initiated by the
strong increase in coupling between non-oscillating electrically
active and passive cells of a pregnant uterus shortly before
delivery. Further justification for this claim stems from the fact
that close contact between ICLCs and smooth muscle cells has been
observed~\cite{Duquette2005}. Moreover, while there has thus far
been no direct evidence for the electrical coupling between myocytes
and fibroblasts via gap junctions, analogous \textit{in vitro} studies
of cardiac tissue~\cite{Kohl2005,Chilton2007} strongly suggest
the occurrence of such coupling. The resting potential of ICLCs,
$V_I^r$ is around $V_I^r \sim - 58 \pm 7 mV$~ \cite{Duquette2005}
and, while the resting potential of fibroblasts, $V_F^r$, varies
over a large range  ($-70$ mV to $0$ mV), it is mostly in the range
$-25mV \le V_F^r \le 0 mV$ (in $77.3\%$ of all cases), with a peak of
the distribution at $-15mV$~\cite{Kiseleva1998}. As both cells
have resting potentials larger than that of the myocyte, they can act
as a source of depolarizing current on coupling, and are thus
potentially significant participants in the generation of spontaneous
activity~\cite{Jacquemet2006,Singh2012}. However, the argument
that spontaneous uterine activity is a result of coupling between
electrically active and passive cells has thus far been tested only
on a highly simplified model of myocyte electrical
activity~\cite{Singh2012}, and no significant attempt has yet
been made to relate the model parameters to actual observations.

Recently developed realistic, biologically detailed
models of uterine myocytes~\cite{Rihana2009,Tong2011} allow for a precise
theoretical study of the roles of individual physiological components
in the generation of desirable, as well as pathological,
electrical activity, which in turn permits a better understanding of
their correlation with contractile
force~\cite{Bursztyn2007a,Maggio2012}. The purpose of the present work
is to investigate the effect of cell coupling on spontaneous
electrically activity using an electrophysiologically realistic
mathematical model, and to examine the synchronization behaviour that
occurs when this coupling is sufficiently strong.
To this end, we present a model for the electrical activity of
uterine smooth muscle cells coupled to passive cells. This
model is based on a realistic mathematical description of uterine
myocyte activity recently developed by Tong~\emph{et
al.}~\cite{Tong2011}, which uses a general Hodgkin-Huxley
formalism to describe the evolution of the membrane potential of
myocytes, $V_{m}$, and the Calcium ionic concentration in the cytosol,
$[\rm{Ca}^{2+}]_{i}$. The details of our model are discussed in
the Methods section. The most significant modification we make
to the model of Tong~{\em et al.} is the addition of an extra current,
arising from the electrical coupling, and an associated equation for
the evolution of the passive cell potential.

In the following section, we present the results of a systematic
investigation into the conditions that give rise to spontaneous
electrical activity, in particular the dependence of myocyte activity
on gap junction conductivity, passive cell resting potential and the
number of passive cells. Furthermore, we examine the regimes that
arise when myocytes and passive cells are coupled in a two-dimensional
($2$-D) assembly. This $2$-D configuration mimics a cell culture of
the type routinely used in cardiac
preparations~\cite{Bub2002,Pumir2005}, or in experiments
performed on the pregnant uteri of small
animals~\cite{Lammers2008}, thus facilitating the potential
experimental verification of our observations.  Our numerical results
strongly suggest that coupling plays an important role in both the
appearance of oscillations as well as in the emergence of synchronized
activity in the tissue. Moreover, we find that our model
is capable of capturing rich dynamical regimes, characterized by
periodically spaced, irregular patterns of action potentials, that are
qualitatively consistent with recent observations
\cite{Lammers2008}.



\section{Materials and methods}
\label{sec:mat_meth}

Our mathematical model builds upon the description of uterine
myocyte activity developed by Tong \emph{et al.}~\cite{Tong2011},
which consists of a set of first order ordinary differential
equations that describe the evolution of fourteen ionic currents,
including depolarizing Na$^{+}$ and Ca$^{2+}$ currents and
repolarizing K$^{+}$ currents. The description of each ionic current
involves activating and inactivating gating variables,
$m_h$ which specify the state of each channel $h$, and are governed 
by evolution equations of the type:
{\begin{equation}
\label{equ:excitable_cell_2}
\frac{\dd m_h}{\dd t} = \frac{m_h^{\infty}-m_h}{\tau_h}\,,
\end{equation}}
where  $m_h^{\infty}(= \alpha_h/(\alpha_h + \beta_h))$ are the asymptotic values of $m_{h}$, $\tau_h (= 1/(\alpha_h + \beta_h))$ are the relaxation times, and $\alpha_h$ ($\beta_h$) are the rates at which the channels open (close). The relaxation times are represented by nontrivial functions of the membrane potential, $V_{m}$, that are typically determined experimentally.

We describe the excitation dynamics of myocytes in terms of the time evolution of this membrane potential:
{\begin{equation}
\label{equ:excitable_cell}
C_m\frac{\dd V_m}{\dd t} = - I_{\rm ion}+I_{\rm ext} + I_{\rm gap}\,,
\end{equation}}
where $C_m$ is the cell membrane capacitance, $I_{\rm ion}$ is the sum
of the fourteen trans-membrane ionic currents and $I_{\rm ext}$
accounts for any externally applied current. This expression
differs from that used in the model by Tong \emph{et
al.}~\cite{Tong2011} in that we include an additional gap-junction
mediated coupling current,
$I_{\rm gap}$. This term accounts for the current $I_{\rm
gap}^{p}$ induced by the interaction of myocytes with passive cells
and, in the case of a $2$-D lattice, the additional inter-myocyte
coupling current $I_{\rm
gap}^{m}$. We use the standard convention where 
outward ionic currents and externally applied currents are taken as
positive.
For the purposes of the present study, we ignore the effect of
external currents and set $I_{\rm ext}=0$.

The model by Tong \emph{et al.}~\cite{Tong2011} also describes the
evolution of the intra-cellular Calcium ion concentration, $\Ca$, in the cytosol,
{\begin{equation}\label{equ:Ca}
\frac{\dd\Ca}{\dd t} = -(J_{\rm Ca,mem} + J_{\rm PMCA} + J_{\rm NaCa}) \,,
\end{equation}}
where the flux of Calcium ions has three components: (i) $J_{\rm
Ca,mem}$, which represents Calcium flux from specific membrane
channels, including L and T-types and other nonspecific cation
currents; (ii) $J_{\rm PMCA}$, which represents the flux of
plasmalemmal Ca$^{2+}$-ATPase; and (iii) $J_{\rm NaCa}$, which
represents flux from Na$^+$-Ca$^{2+}$ exchangers. The currents
resulting from the plasmalemma and from the exchangers both extrude
Calcium ions from the cell. In particular, the Na$^+$-Ca$^{2+}$
exchangers extract one Ca$^{2+}$ ion from the cytosol for three
Na$^+$ ions pumped into the cell \cite{Weber2001}. By their very nature,
the ionic currents $J_{\rm PMCA}$ and $J_{\rm NaCa}$ must
extrude Calcium, and as such, must be positive. Consequently, the
current resulting from the action of the exchanger, $I_{\rm NaCa}$, is
inward (or repolarizing) and hence, by the standard convention,
negative. 
In our model we have 
ensured that the known physiological functions of the exchangers are
cogently described, and that the
requirements that $J_{\rm NaCa} \ge 0$ and $I_{\rm NaCa} \le 0$
are satisfied.
Additionally, motivated by the experimental literature (in
particular~\cite{Weber2001}) as discussed in Sec.~S.2 of the SI,
we used different parameter values for the terms that describe the
Na$^+$-Ca$^{2+}$ exchanger. 

In order to motivate the electrical coupling mediated by
gap junctions between
myocytes and passive cells, we note the
observation~\cite{Duquette2005} that although ICLCs do not exhibit
regular spontaneous depolarizations and appear unable to generate
action potentials, the application of an external current
causes their membrane potential to relax at a
near-exponential rate with a characteristic time scale of $\sim
0.2-1s$
(see Fig.7B of ~\cite{Duquette2005}). Thus, when a passive cell
of this type is electrically coupled to a myocyte, its membrane
potential dynamics can be described by:
{\begin{equation}\label{eq:coup_ICLC}
C_P\frac{\dd V_P}{\dd t} = {{G^{\rm int}_P}}(V_P^r - V_P) + I^p_{\rm gap}\,,
\end{equation}}
where $C_P$, ${G^{\rm int}_P}$ and $V_P^r$ represent the capacitance,
conductance and resting  potential, respectively, of a generic passive
cell and $I^p_{\rm gap}$ is the coupling current.
As a consequence, a myocyte with $n_{p}$ passive cells in its
neighborhood experiences a coupling current $I_{\rm gap} = -
n_{p}\, I^p_{\rm gap}$. The current $I^p_{\rm gap}$ is
proportional to the difference between the potentials of the myocyte,
$V_m$, and passive cell, $V_P$, across the electrically conducting
pores that result from the existence of gap junctions. The
coupling-induced current can hence be expressed as $I^p_{\rm gap} =
G_p\,(V_m - V_P)$, where the gap junction conductance, $G_p$, is
directly related to the level of expression of the connexin proteins
that constitute these junctions. As the conductance of a single gap
junction channel has been estimated to be of the order of 50
pS~\cite{Valiunas2004}, the relation between the conductance and
then
number of expressed gap junctions $n_{gj}$ is simply $n_{gj} \approx
G_p/50$~pS.

We note that although Eq.~(\ref{eq:coup_ICLC}) is sufficient for
our current purposes, it does not capture the full complexity of the
passive cell membrane dynamics. As seen in Fig.7B
of~\cite{Duquette2005}, when the applied current is varied, the
relaxation time scale changes, suggesting a dependence of {$G^{\rm int}_P$} as a
function of the membrane potential. We further note that although
Eq.~(\ref{eq:coup_ICLC}) was formulated based on known properties of
ICLCs, it can also be used to describe the behaviour of other
electrically passive cells, such as fibroblasts, which have a membrane
conductance of $G^{\rm int}_F=1nS$~\cite{Kohl1994}. Indeed, it is
instructive to consider the case where myocytes are
simultaneously coupled to different types of passive cells (see
Sec.~S4 of the SI for more details).

\begin{figure}[h] 
   \centering
   {\includegraphics[width=10.0cm]{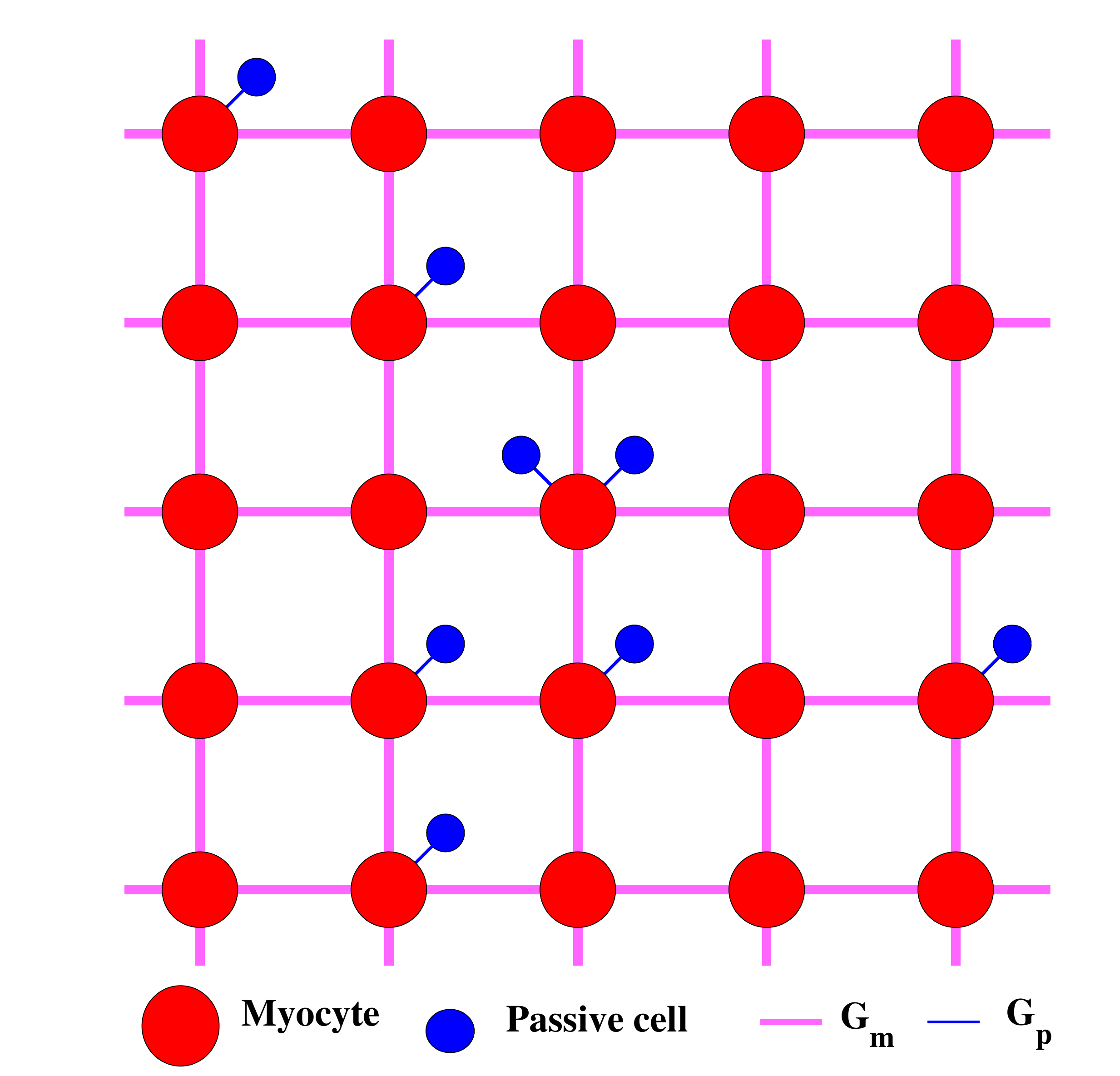} }
   \caption{Schematic representation of the $2$-D square lattice of
   uterine myocytes (shown in red), each myocyte coupled to a random number of
   passive cells (shown in blue). Neighboring myocytes are
   electrically coupled with strength $G_m$ and the coupling strength
   between a myocyte and a passive cell is $G_p$.
}
   \label{fig:schematic}
\end{figure}

In order to describe the effect of inter-myocyte coupling, we
assume that myocytes are coupled to their nearest neighbors on a
$2$-D square lattice of size $N\times N$, and label each cell by the
indices of its row ($a$) and column ($b$). In this case, each myocyte
receives a coupling current $I^{m}_{\rm gap}$ given by
{\begin{equation}
I^{m}_{\rm gap}(a,b) = G_m \Bigl(V_m(a+1,b) + V_m(a-1,b) + V_m(a,b+1) + V_m(a,b-1) - 4 V_m(a,b) \Bigr)\,,
\label{equ:coupl}
\end{equation}}
where $G_m$ is the conductance of the myocyte gap junctions. The
total coupling current experienced by a myocyte coupled to both
electrically passive cells, as well as other myocyte cells in a
lattice (see Fig.~\ref{fig:schematic}) is thus $I_{\rm gap} = -
n_{p}\, I^{p}_{\rm gap} + I^{m}_{\rm gap}$. The coupling
current Eq.~(\ref{equ:coupl}) has the form of a diffusive term,
with
$G_m/C_m$ acting as an effective diffusion constant. 

Finally, we note that one of the relaxation times
$\tau_{\alpha}$ can be of the order of several hundred pico seconds,
and hence, the maximum simulation time step is greatly
constrained. However, this very short time scale implies that
the gating variable $x_\alpha$ relaxes very quickly to its
steady-state value. Thus, in order to increase numerical
efficiency, we make the assumption that $x_{\alpha}(t) =
x_{\alpha\infty}(V_m)$, which allows us to use a comparatively
larger time step $dt=0.5ms$. We have verified that the numerical
results are not sensitive to this approximation, with only a $0.25\%$
error in the periods of oscillation.

Apart from the details described above, our model equations and
the parameter values used in our simulations correspond
exactly to those used in Ref.~\cite{Tong2011}. Our numerical simulations
involved the integration of the set of ordinary differential
equations using a standard fourth-order Runge-Kutta scheme. For
improved efficiency, the Message-Passing-Interface (MPI) library was
implemented to distribute the load among up to 16 processors. When
obtaining time series data, care was taken to ensure that any
transient behaviour was discarded.
While our model behaves slightly differently from the model by
Tong \emph{et al.}~\cite{Tong2011} in response to external stimuli
(see Sec.~S.3.2 of the SI), we emphasize that, as in
Ref.~\cite{Tong2011}, results obtained using our model are consistent
with data from voltage clamp experiments (see Sec.~S.3.1 of the SI
for details of our validation tests).
In addition, we have verified that our results are qualitatively
robust, by simulating both the single cell and $2$-D cases with
slightly different sets of parameters, different realizations
and, in the $2$-D case, a larger lattice.
We note that
for a sufficiently large lattice size, the results do not
qualitatively depend on $N$.


\section{Results}

We now investigate the hypothesis mentioned in the introduction,
namely that the
interaction between coupled myocytes and passive cells is fundamental
to their spontaneous activation during the late stages of pregnancy.
As the number of gap junctions are known to increase during this
period~\cite{Miyoshi1996}, one expects higher values for the effective coupling conductances $G_p$
and $G_m$. Moreover, although individual passive
cell types are each characterized by a unique resting potential, a
mixture of passive cell types can result in an effective $V_p^r$ that
is different from those of the constituent cells. The effect of such
phenomena on the dynamical behavior of the coupled system are shown
below. In the following subsections, we
present a numerical investigation
of the electrical activity of myocyte cells coupled to $n_p$ passive
cells, followed by a study of the emergence of regimes of
regular and irregular dynamical activity in a $2$-D lattice of
myocytes coupled to each other, as well as to passive cells.

\subsection{Coupling a single myocyte to electrically passive cells}

A myocyte is known to exhibit oscillatory behaviour when
external current is injected into it (see SI for
more details). As coupling through gap junctions with
neighboring cells provides a source of such an inward current,
the electrical state of a myocyte coupled
to passive cells can be dynamically modulated by changing the
strength of this coupling.
When neighboring myocytes in a tissue are
strongly coupled, as would be expected towards the
late stages of {pregnancy~\cite{Miyoshi1996},} their behaviour is sensitive to
the average number of passive cells in the tissue -- a property that
we have explicitly verified through numerical simulations on an
assembly of cells~\cite{Xu2013}. The limiting case of large
coupling between myocytes can be approximated by considering a single
myocyte coupled to $n_{p}$ passive cells. Note that as $n_{p}$ in this
case effectively corresponds to the average number of passive cells in
the tissue, it can take non-integer values.
In the following, we investigate the dependence of the myocyte
behaviour on the conductance of gap junctions between myocytes
and passive cells, $G_p$, the passive cell resting potential,
$V_{p}^{r}$, and the average number of passive cells coupled to
a myocyte, $n_{p}$. 

In our simulations, we have assumed that the myocyte membrane
capacitance is $C_m=120 pF$ and the passive cell has capacitance
$C_P=80 pF$ and conductance {$G^{\rm int}_P=1.0nS$.}
This is motivated by observations that the capacitance of
rat myocytes is around $120 pF$ during the late stages of
pregnancy~\cite{Yoshino1997}, and that the capacitance and input
resistance
of an ICLC are $84.8\pm 18.1 pF$ and $3.04\pm 0.5 G\Omega$,
respectively~\cite{Duquette2005}.
Additionally, we assume that the sodium conductance is
$g_{Na} = 0.04 nS/pF$, which is in the range of values measured during
late pregnancy (see \cite{Tong2011} and references therein). We
have verified that our results are robust with respect to small
changes in the system parameters.

\subsubsection{Dependence on the gap junction conductance}

\begin{figure}[h]
\begin{center}
\subfigure[]{
	\includegraphics[width=7.0cm]{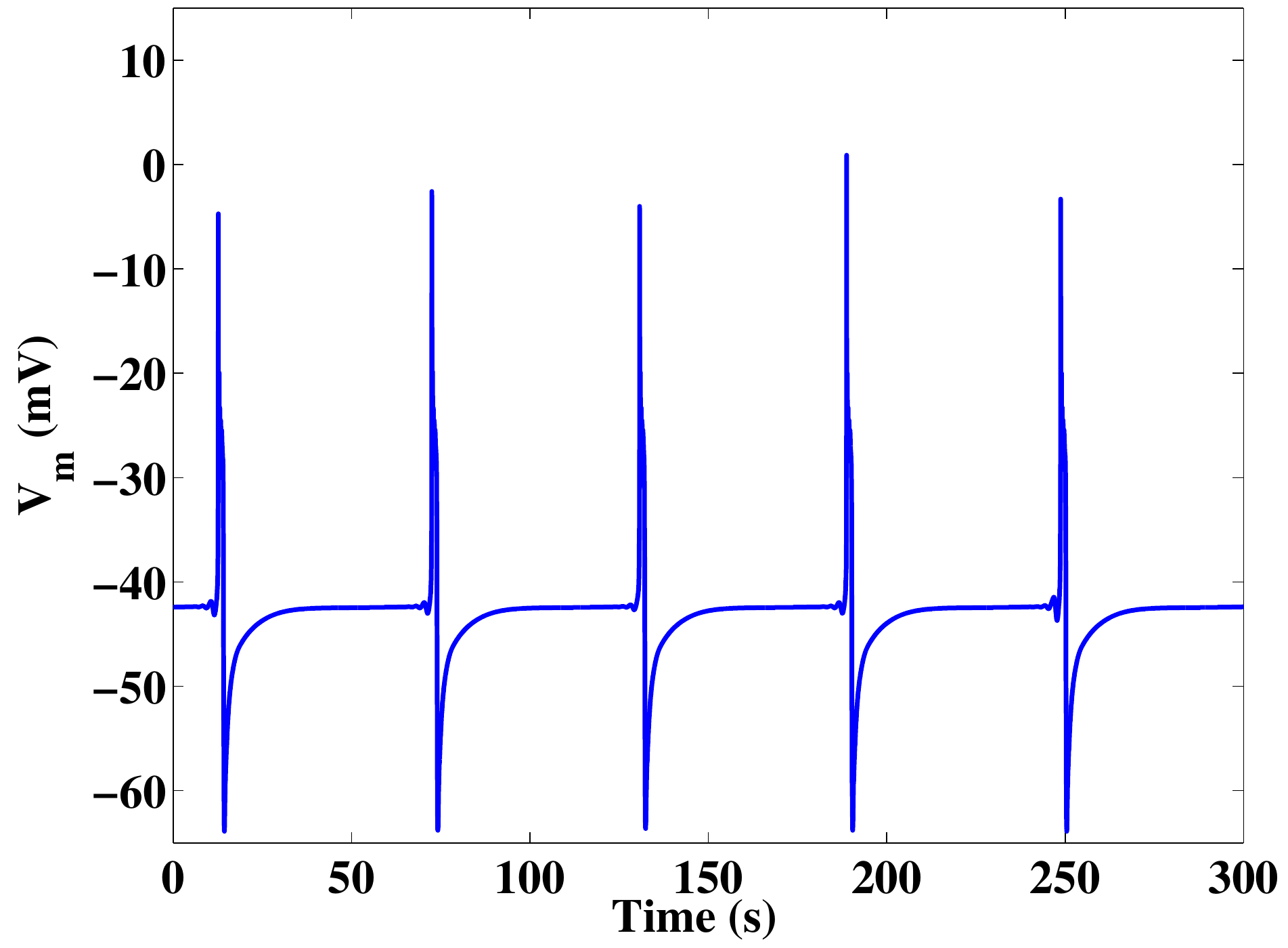}
}
\subfigure[]{
	\includegraphics[width=7.0cm]{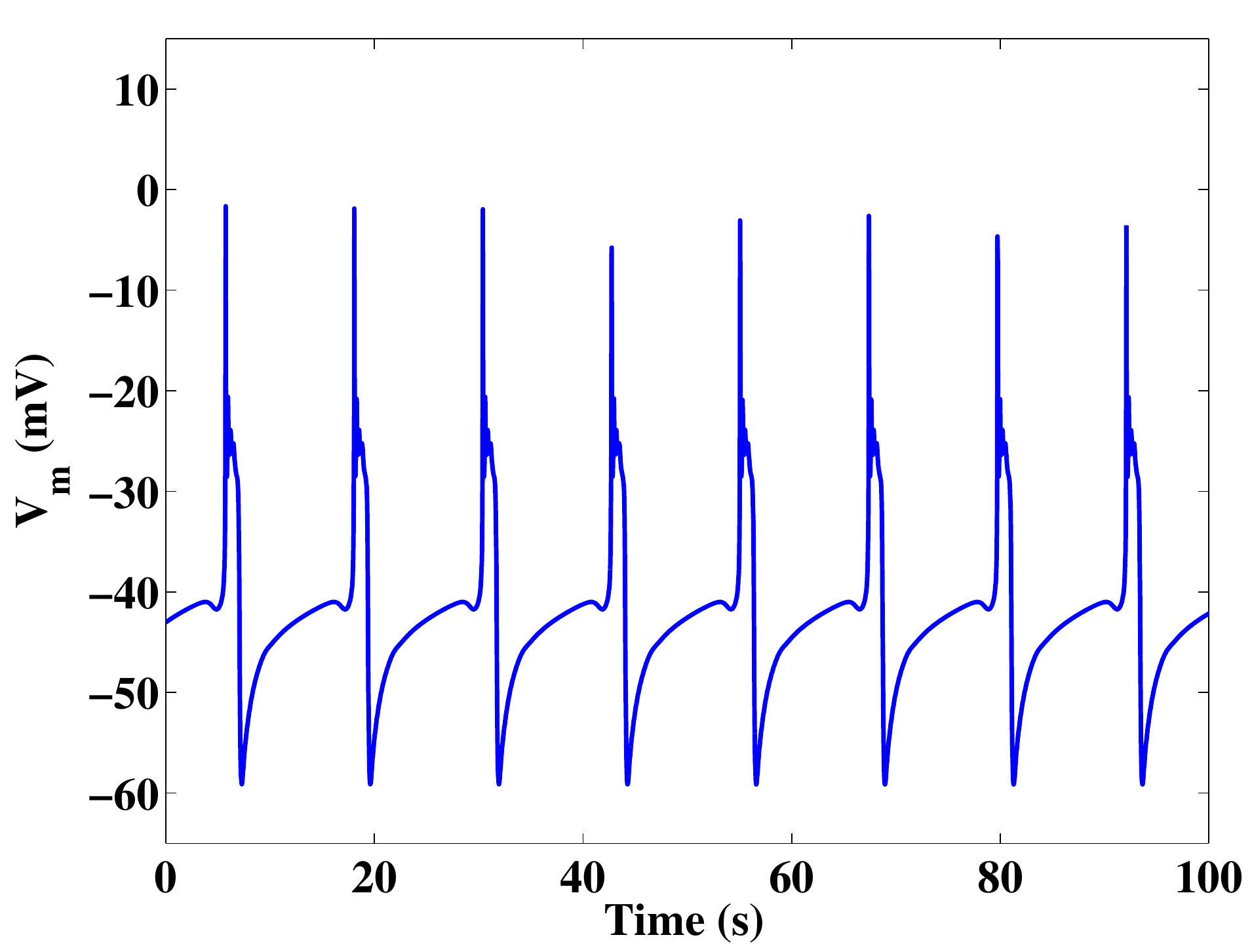}
}
\subfigure[]{
	\includegraphics[width=7.0cm]{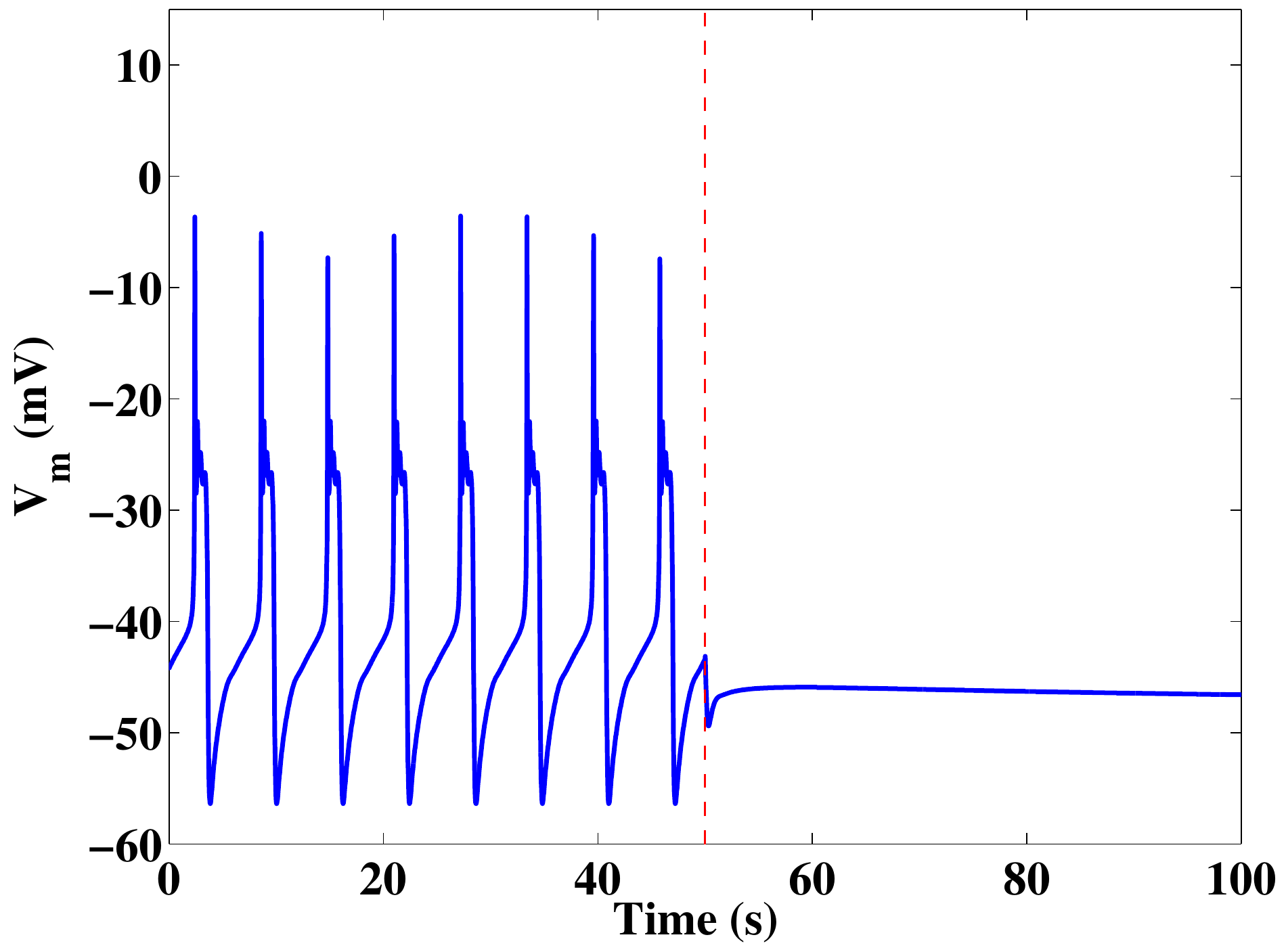}
}
\subfigure[]{
	\includegraphics[width=7.0cm]{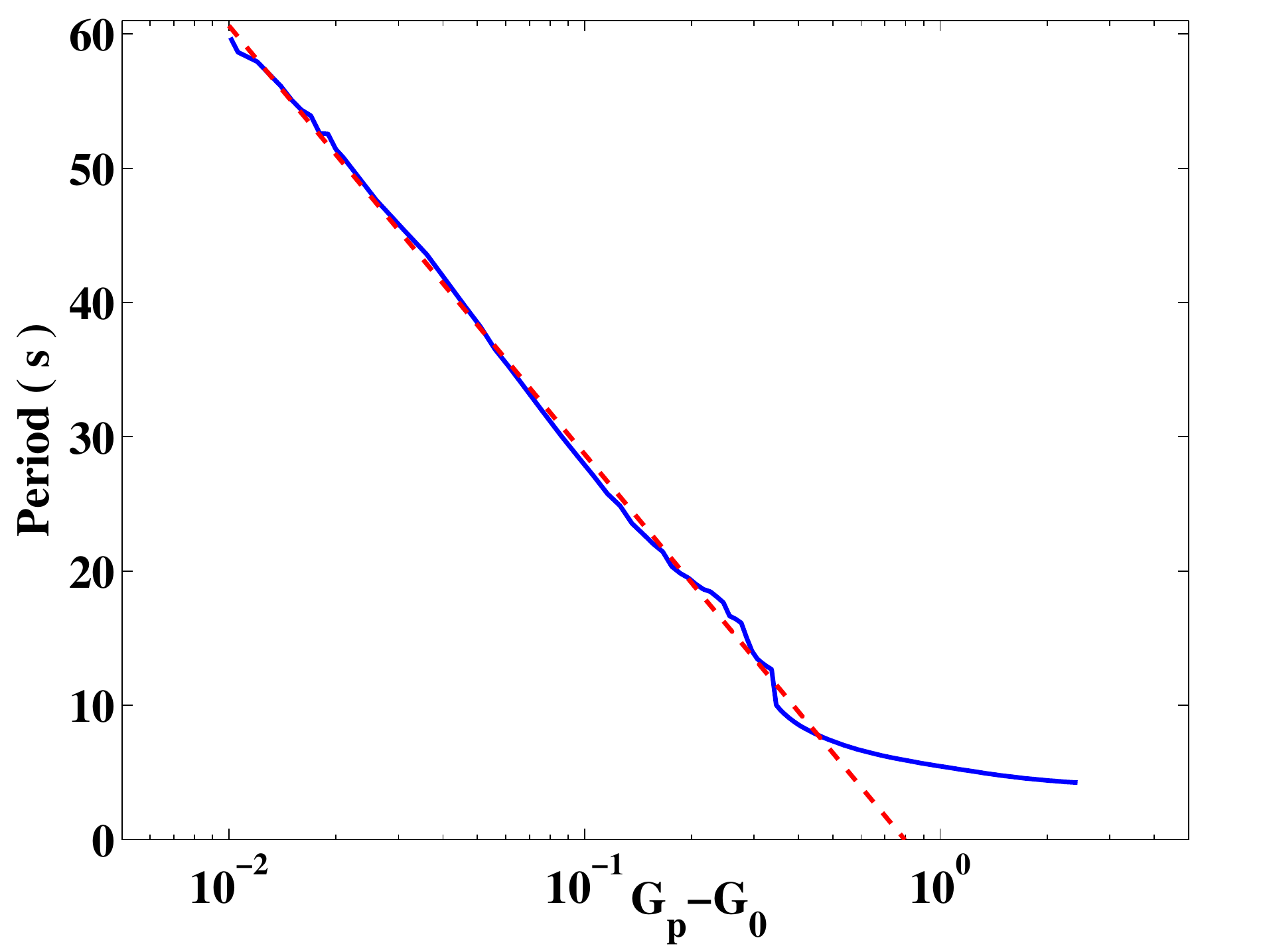}
}
\caption{Dynamical behaviour of a myocyte coupled to a passive cell
for different values of the gap junction conductance $G_p$, and
with $V_p^r = -35 mV$ and $n_p = 1$. (a) At $G_p=0.174 nS$, the
period is $\sim$ 1min. (b) At $G_p=0.5nS$, the period has reduced
to $\sim 12s$. (c) At $G_p= 1 nS$, the period reduces further to
$\sim 7 s$. Here we observe that the sudden uncoupling of the
myocyte and passive cell, at the time indicated by the vertical
broken line, immediately terminates activity. (d) Above a critical value
$G_0 = 0.164 nS$, we observe a logarithmic divergence in the
oscillatory time period: $T \sim \log[(G_p-G_0)/G_0]$ as indicated
by the broken line.
}
   \label{fig:osc_vs_G}
\end{center}
\end{figure}

The evolution of the membrane potential of a myocyte, $V_{m}$,
coupled to a single passive cell is displayed in
Fig.~\ref{fig:osc_vs_G} for different values of the gap junction
conductance, $G_p$. For
values of $G_p$ less than a critical threshold $G_{0}\approx 0.164
nS$, $V_{m}$ approaches a steady state, while for values of
$G_p>G_{0}$, the electrical coupling induces spontaneous temporal
oscillations whose time period decreases as $G_p$ increases. As seen
in Fig.~\ref{fig:osc_vs_G}a, the oscillation time period for $G_p =
0.1741 nS$ is $T\sim 1$ min,  while for $G_p = 0.5 nS$, we find $T
\sim 12 s$ (Fig.~\ref{fig:osc_vs_G}b), and for $G_p = 1 nS$, we find
$T \sim 7s$  (Fig.~\ref{fig:osc_vs_G}c). As shown in
Fig.~\ref{fig:osc_vs_G}c, the oscillatory behaviour can be
suppressed immediately upon uncoupling the myocyte and the passive
cell, i.e., by setting $G_p$ to $0$. This is consistent with the
experimental observation that the addition of a gap junction uncoupler
leads to rapid termination of electrical
activity~\cite{Tsai1998}. As $G_p$ approaches $G_0$ from above, the
time period $T$ grows like $T \sim \log[(G_p-G_0)/G_0]$ (see
Fig.~\ref{fig:osc_vs_G}d). Although this logarithmic divergence of 
time periods close to a critical point is suggestive of a homoclinic
bifurcation~\cite{GuckHolm1983}, we note that the dynamical behaviour
in the interface between the regimes of activity and inactivity is
in fact more complicated. We observe that 
there exists a
small range of values of $G_p$ for which both stable and oscillatory
solutions are possible. Depending on the precise choice of initial
condition, the system may evolve to either of the two asymptotic
solutions (attractors), corresponding to quiescence or oscillations.
%

\subsubsection{Dependence on the passive cell resting potential}

\begin{figure}[h]
   \centering
{\includegraphics[width=8.0cm]{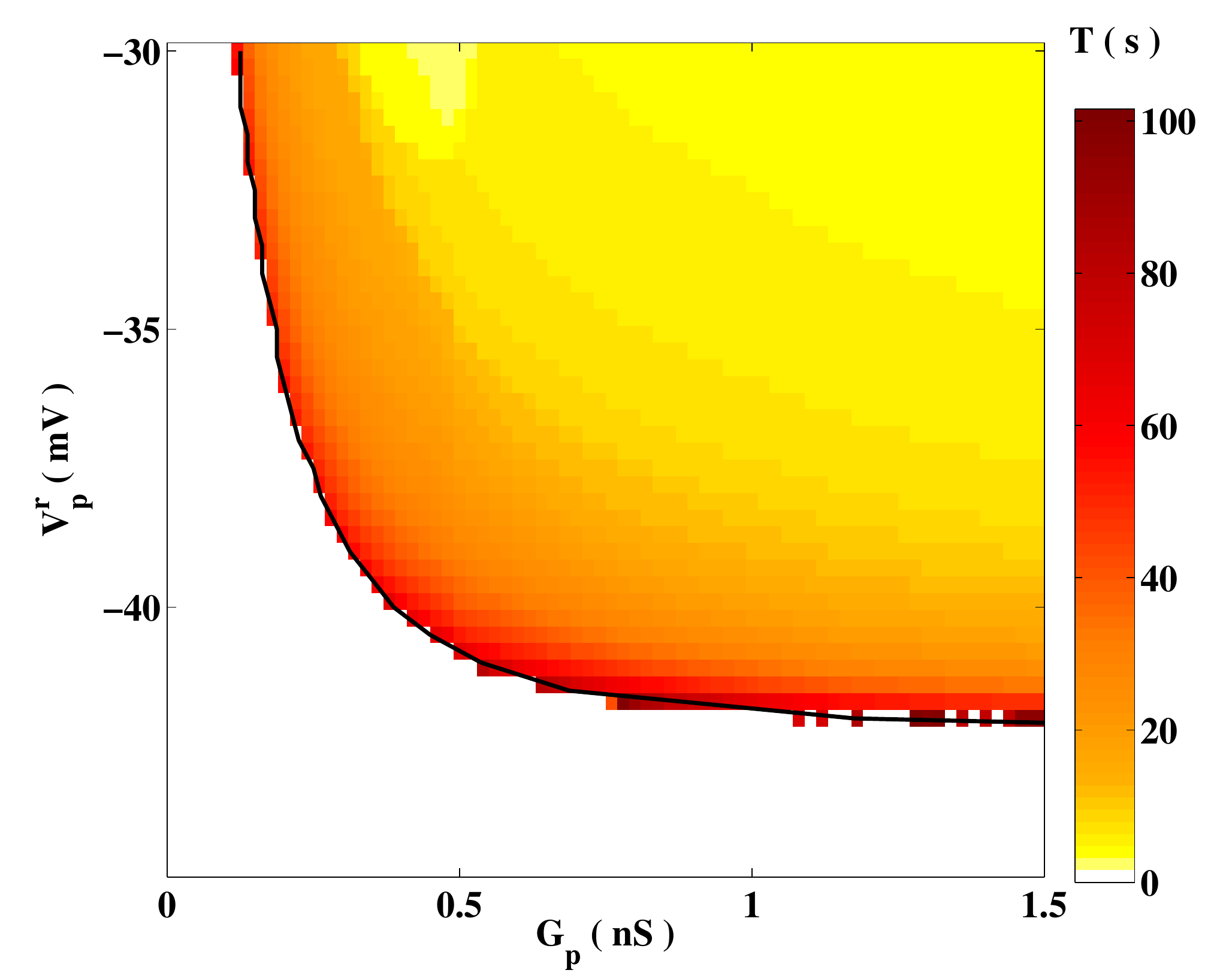}}
\caption{Time periods $T$ (in sec) of the oscillation of the
membrane potential, for a range of values of $V_p^r$ and $G_p$.
Regions shown in white correspond to the absence of oscillatory
activity. The bifurcation from oscillatory activity to a quiescent dynamical
regime occurs at the interface indicated by the continuous 
curve obtained by fitting numerical data.
}
\label{fig:osc_G_Vpr}
\end{figure}

The time periods, $T$, of the oscillations of the
membrane potential, obtained for different values of the passive cell
resting potential, $V_p^r$, and the gap junction conductance, $G_p$,
are
displayed in Fig.~\ref{fig:osc_G_Vpr}. Consistent with the
observations of Fig.~\ref{fig:osc_vs_G}d, the period of oscillation is
very large close to a threshold value of $G_p$ for values of
$V_{p}^{r}$ larger than $\sim - 42 mV$. In addition, $T$ increases
on decreasing $V_{p}^{r}$. For any given $G_p$, there exists
a threshold value of $V_{p}^{r}$ below which the solution approaches a
steady, non-oscillating state. Conversely, as $V_{p}^{r}$ increases,
one finds that $T$ decreases, and can be as low as a few seconds
for large values of $V_p^r$ and $G_p$. As in the situation
discussed in the previous subsection, the system can evolve to either
a quiescent or oscillatory solution when $V_{p}^{r}$ and $G_p$ are
close to the bifurcation line (indicated in Fig.~\ref{fig:osc_G_Vpr}
by a continuous curve).

Fig.~\ref{fig:osc_G_Vpr} suggests, in particular, that the gap
junction conductance threshold is a decreasing function of $V_p^r$.
This can be qualitatively understood by noticing that
coupling a myocyte to a passive cell is equivalent to the addition of a
current $ I_{ext} \sim - \frac{{G^{\rm int}_P} G_p}{{G^{\rm int}_P} + G_p} V_p^r $ in the
expression for myocyte membrane potential,
Eq.~(\ref{equ:excitable_cell}), at least in the
limit of large passive cell relaxation time.
From this expression for the external current $I_{ext}$, elementary
algebraic considerations show that as the potential $V_{p}^{r}$ increases,
the coupling $G_p$ necessary to deliver a given current decreases 
(see also Supplementary Information). 

%
\subsubsection{Dependence on the number of passive cells}

\begin{figure}[h] 
   \centering
{\includegraphics[width=16.0cm]{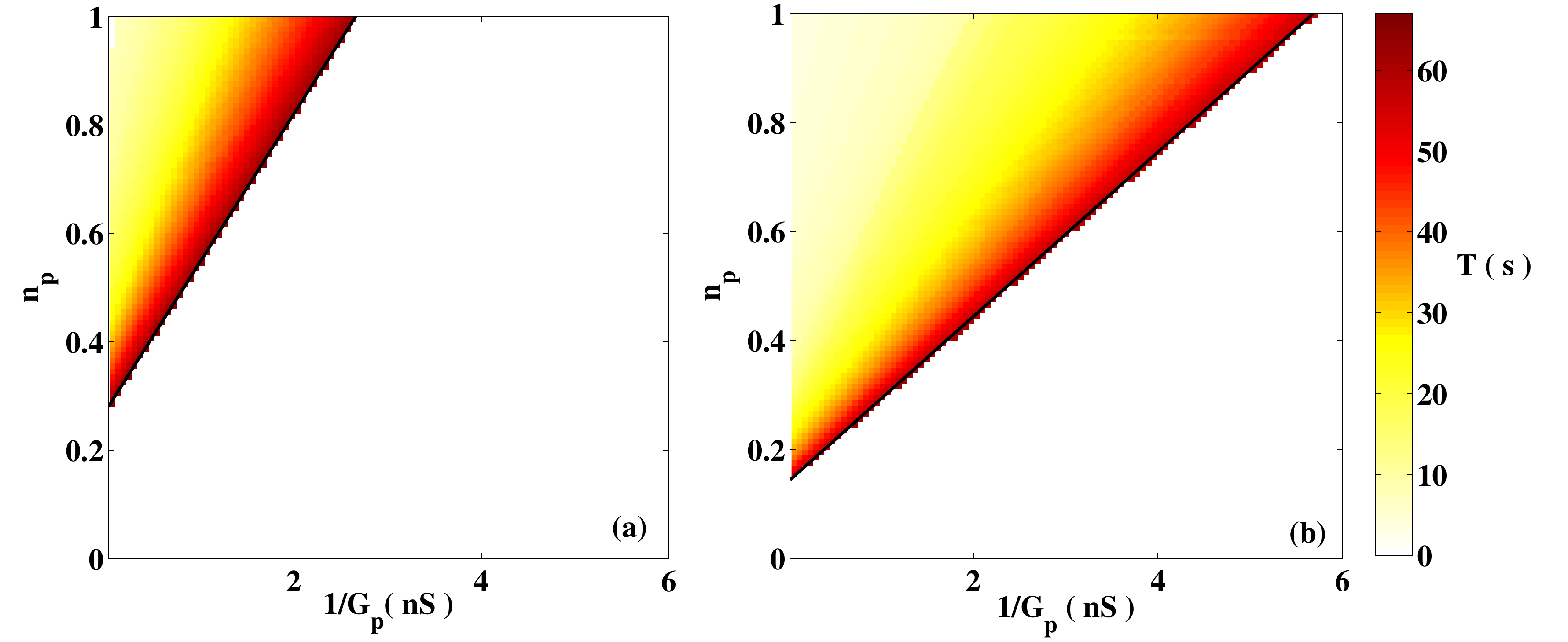}}
    \caption{ Time periods of oscillation, $T$ (in sec) at two
    different values of $V_p^r$, for a range
    of values of $n_{p}$ and $G_p$. Regions shown in white
    correspond to the absence of oscillatory activity.
    Activity is seen when $n_p \ge A
    + B/G_{p}$ (indicated by the solid line), a functional form that can be justified from
    elementary considerations, see text. (a) $V_p^r = -40 mV$
    {($ A \approx 0.28$, $B \approx 0.27$)} (b) $V_p^r = -35 mV$ 
{($A \approx 0.17$, $B \approx 0.15$)}.
}
    \label{fig:osc_G_vs_np}
\end{figure}

It is known that the total number of passive cells in
uterine tissue is only a fraction of that of the
myocytes~\cite{Young2007}. As mentioned earlier, 
each myocyte is attached to $n_p$ passive cells.
Fig.~\ref{fig:osc_G_vs_np} displays the domain
of oscillatory activity in the $( G_p, n_p)$ plane, observed for two
different choices of the passive cell resting potential: $V_p^r = - 40
mV$ (Fig.~\ref{fig:osc_G_vs_np}a) and $V_p^r = -35 mV$
(Fig.~\ref{fig:osc_G_vs_np}b). In each case, we find that the cell is
quiescent for low values of $G_p$ and $n_p$, while spontaneous
oscillatory activity is generated on increasing these parameters. We
note from Fig.~\ref{fig:osc_G_vs_np} that for any given value of
$G_{p}$ there exists a critical value of $n_{p}$ below which
spontaneous oscillations will not occur. In the limit
$1/G_p\rightarrow 0$, we find this critical value to be $n_p \approx
0.28$ ($n_p \approx 0.17$) for a passive cell resting potential $V_p^r
= -40 mV$ ($V_p^r = -35 mV$).

We observe that the curve delimiting the region of activity in
Fig.~\ref{fig:osc_G_vs_np} has a simple analytic form $ n_p = A/G_p +
B$, where the parameters $A$ and $B$ are related by $B = A/G^{\rm int}_P$. 
In the limit where the passive cell resistivity $G^{\rm int}_P$ is
large, the passive cell membrane potential relaxes
quickly to its equilibrium value. 
From Eqs.~(\ref{equ:excitable_cell}) and
(\ref{eq:coup_ICLC})
one can show that the coupling term
in Eq.~(\ref{equ:excitable_cell}) is equivalent to adding an external
current
$ I_{ext} \sim - n_p \frac{{G^{\rm int}_P} G_p}{{G^{\rm int}_P} + G_p} V_p^r = - n_p/ \left(\frac{1}{{G^{\rm int}_P}} + \frac{1}{G_p} \right) V_p^r $.
This expression implies that the effect of the coupling in the
equation for the myocyte action potential
depends on the quantity $n_p/ \left(\frac{1}{G^{int_P}} + \frac{1}{G_p}
\right)$, which is constant provided $n_p $ is proportional to 
$1/G^{\rm int}_P + 1/G_p$. 
Hence the analytical form for $n_p$ stated above delineates the region of
parameter space where oscillations occur
(Fig.~\ref{fig:osc_G_vs_np}).


\subsection{Coupling myocytes to passive cells in a $2$-D lattice}
\label{sub:2D}


We now investigate dynamical patterns of activity that arise
on a spatially extended domain characterized by nearest-neighbor
interactions between the myocytes.
To this end, we
assume that myocytes are electrically coupled in an $N\times N$
square lattice (Fig.~\ref{fig:schematic}); we take $N = 50$ in the present 
study.
Additionally,
each myocyte is coupled to an integer number of
passive cells $n_p$ drawn from a random distribution $P(n_p)$,
chosen here to be
binomial with mean $f=<n_{p}>=0.2$.
Numerical simulations reveal that the precise choice of boundary
conditions for our model can affect some qualitative features of the
observed patterns. In this paper, we limit our $2$-D investigation to
the study of the dynamics in an isolated segment of late-pregnant
myometrium by imposing no-flux boundary conditions on $V_{m}$, thus
allowing for direct comparisons with experiments, such as those
performed by Lammers and coworkers~\cite{Lammers1997,Lammers1999,Lammers2008}.
In the following, we set $G_p = 3.5nS$ and systematically vary
the inter-myocyte gap junction conductance $G_m$.

{\subsubsection{Dependence on the inter-myocyte gap junction conductance}
\label{subsub:regular}}

The dependence of the electrical activity of the lattice on the
inter-myocyte gap junction conductance $G_m$ is shown in
Fig.~\ref{fig:space_activity}. As the coupling strength is known to
increase during pregnancy {\cite{Miyoshi1996}}, these results are plausibly
indicative of the transition towards coherent activity in the uterus
close to term.

For low values of $G_m$, a significant number of cells remain
quiescent, while the remaining cells organize themselves into a few
localized clusters, each of which are characterized by a unique
oscillating frequency. This regime, referred to as cluster
synchronization (CS), is shown in Fig.~\ref{fig:space_activity}
(left column), where different clusters with characteristic
effective oscillatory periods in the range $5-10$ sec are observed for $G_m =
0.48nS$.
It has been shown, using a simplified model of myocyte
activity~\cite{Xu2013}, that the electrical activity originates in
regions where the coarse-grained density of passive cells attached to
the myocytes is high. 

On increasing $G_m$, we observe that the various clusters begin to
merge, eventually giving rise to a scenario where all the cells that
oscillate do so with the same frequency (see the middle column of
Fig.~\ref{fig:space_activity}, which displays results for $G_m = 1.8
nS$). In this regime, referred to as local synchronization (LS), it is
appropriate to define the fraction of oscillating cells, by $n_{osc} =
N_{osc}/N^2$ where $N_{\rm osc}$ is the number of oscillating
cells~\cite{Singh2012}. We find the existence of pockets of cells that
remain quiescent, and also observe that regions with a high
coarse-grained density of passive cells produce travelling waves
that propagate through the system, thus effectively acting
as ``pacemaker regions"~\cite{Xu2013}.

As we increase $G_m$ further, we find that every cell in the system
oscillates at exactly the same frequency (see the right column
of Fig.~\ref{fig:space_activity}, which displays results for $G_m =
2.4 nS$). This dynamical state
is known as global synchronization (GS). The
transition from LS to GS is characterized by an increase
in the fraction of oscillating cells $n_{osc}$ to 1.

%
{\begin{figure}[tb] 
  \centering
{\includegraphics[width=12.0cm]{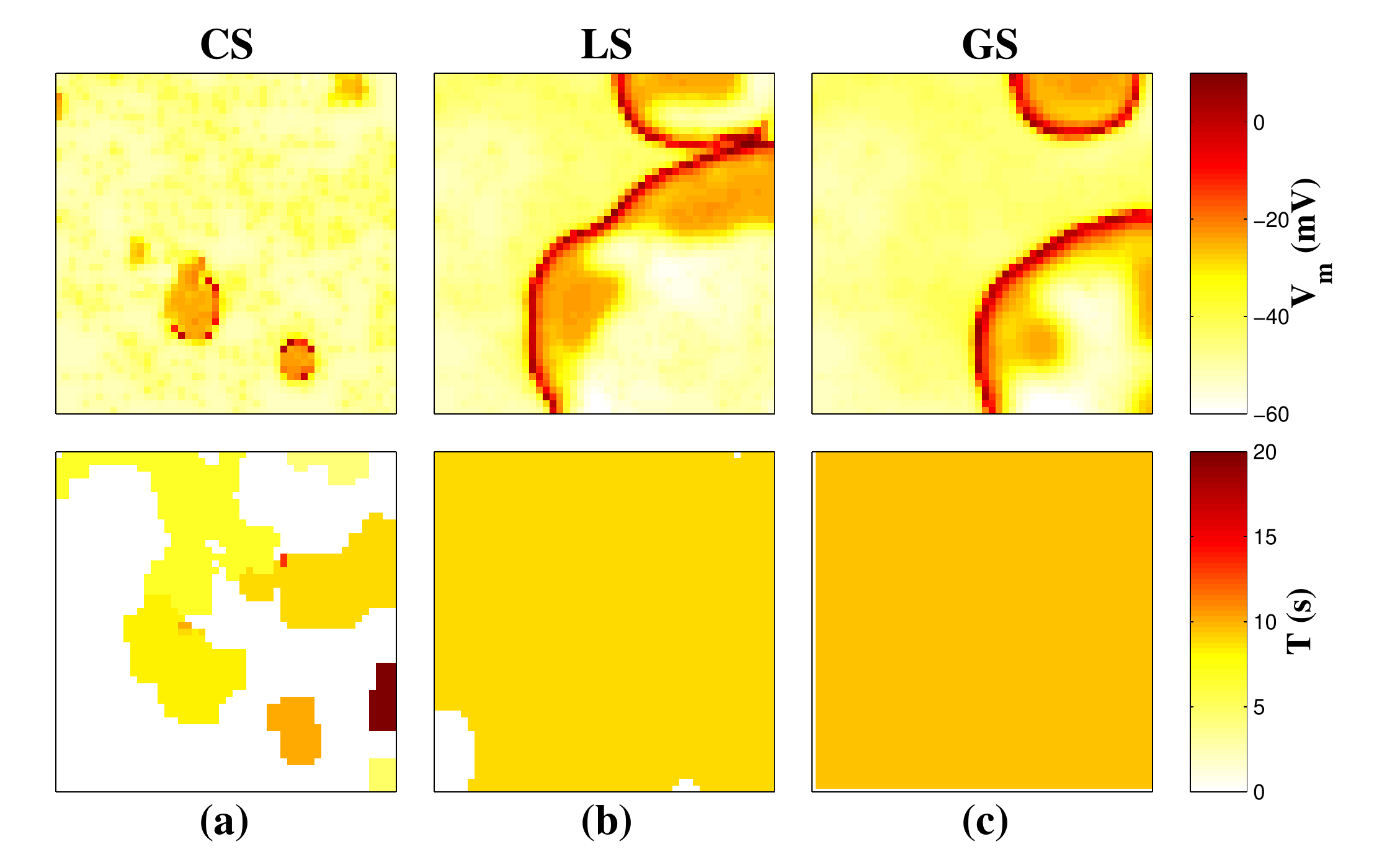}}
  \caption{Patterns of electrical activity observed for different
  coupling strengths $G_m$ between myocytes on a $50\times 50$
  lattice, where each myocyte interacts on average with $f (=0.2)$
  passive cells.
  The upper row shows snapshots of the membrane potential while the
  lower row shows the corresponding effective time periods of
  oscillatory activity. 
  (a) Cluster
  Synchronization (CS) is observed at
  $G_m=0.48 nS$ where cells group into several synchronously
  oscillating clusters, each characterized by a
  different frequency, coexist with regions in which the tissue is at
  rest. (b) At $G_m=1.8 nS$  all cells in the lattice that
  oscillate do so with a single frequency. However, we also observe a few
  non-oscillating cells indicating that this corresponds to the LS
  regime. (c)  At $G_m=2.4 nS$, which lies in the GS regime,
  every cell in the lattice oscillates with the same frequency.
}
  \label{fig:space_activity}
\end{figure}}

The phase diagram in Fig.~\ref{fig:phase_diagram}, obtained over a
range of values of $G_m$ and $G_p$, displays the aforementioned
regimes of dynamical behaviour, viz., CS, LS and GS. 
An additional state is seen for sufficiently low $G_p$ and
sufficiently large $G_m$ where no oscillations (NO) are observed. 
We observe in Fig.~\ref{fig:phase_diagram} that when $G_m$ is
large, the transition between NO and GS occurs at a value of
$G_p \approx 2.7 nS$, consistent with Fig.~\ref{fig:osc_G_vs_np}. 
The phase diagram in Fig.~\ref{fig:phase_diagram} is qualitatively similar to
that obtained with the 
simpler FitzHugh-Nagumo model, used for describing the dynamics
of an excitable cell~\cite{Singh2012}. We note that for a range
of parameter values in the domain corresponding to GS in
Fig.~\ref{fig:phase_diagram}, the activity is in fact irregular,
an issue discussed in Sec.~\ref{subsub:irregular}.

{\begin{figure}[ht] 
    \centering
{\includegraphics[width=12.0cm]{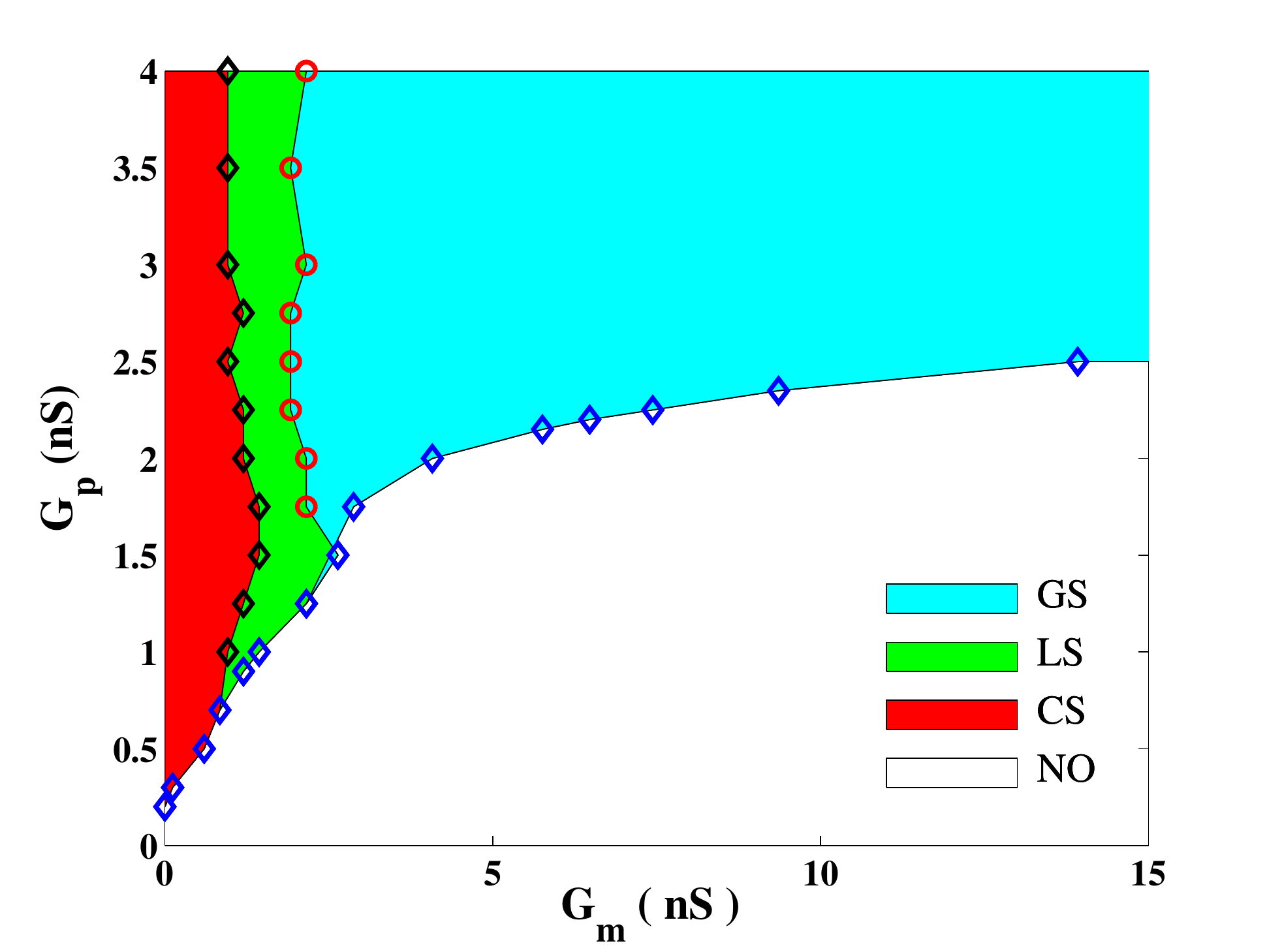}}
\caption{Dynamical regimes observed in the 2-D lattice of
    coupled myocytes and passive cells for
    a range of coupling strengths. Three distinct synchronization
    regimes are observed: CS at low $G_m$,
    LS at intermediate $G_m$ and GS at high
    $G_m$. For every value of
    $G_m$, there exists a critical value of $G_p$ below which no
    oscillations (NO) are observed. 
    The symbols indicate numerically determined points 
    lying on the boundaries between
    the various dynamical regimes.
    }
    \label{fig:phase_diagram}
 \end{figure}}

For a sufficiently large inter-myocyte gap junction conductance
($G_m \gtrsim 19.2 nS$) the activity of the system is characterized
by a simple, regular spatial pattern
(Fig.~\ref{fig:periodic_activity}). Here, an action potential is emitted
in the form of a target wave from a single, dominant region
of high passive cell density, which effectively acts as a local
pacemaker. The nature of the transition from distinct, competing
wave sources at low $G_m$ to a single dominant source at large $G_m$
is explicated in Ref.~\cite{Xu2013}. The range of values
of $G_m$ spanned in Fig.~\ref{fig:phase_diagram} correspond
to experimentally relevant values of inter-cellular
coupling in the uterine myometrium~\cite{Miyoshi1996}. 
Based on the results of Ref.~\cite{Singh2012},
we expect that coherent activity over the entire system will be
observed at even larger values of $G_m$.

{\begin{figure}[ht] 
   \centering
{\includegraphics[width=12.0cm]{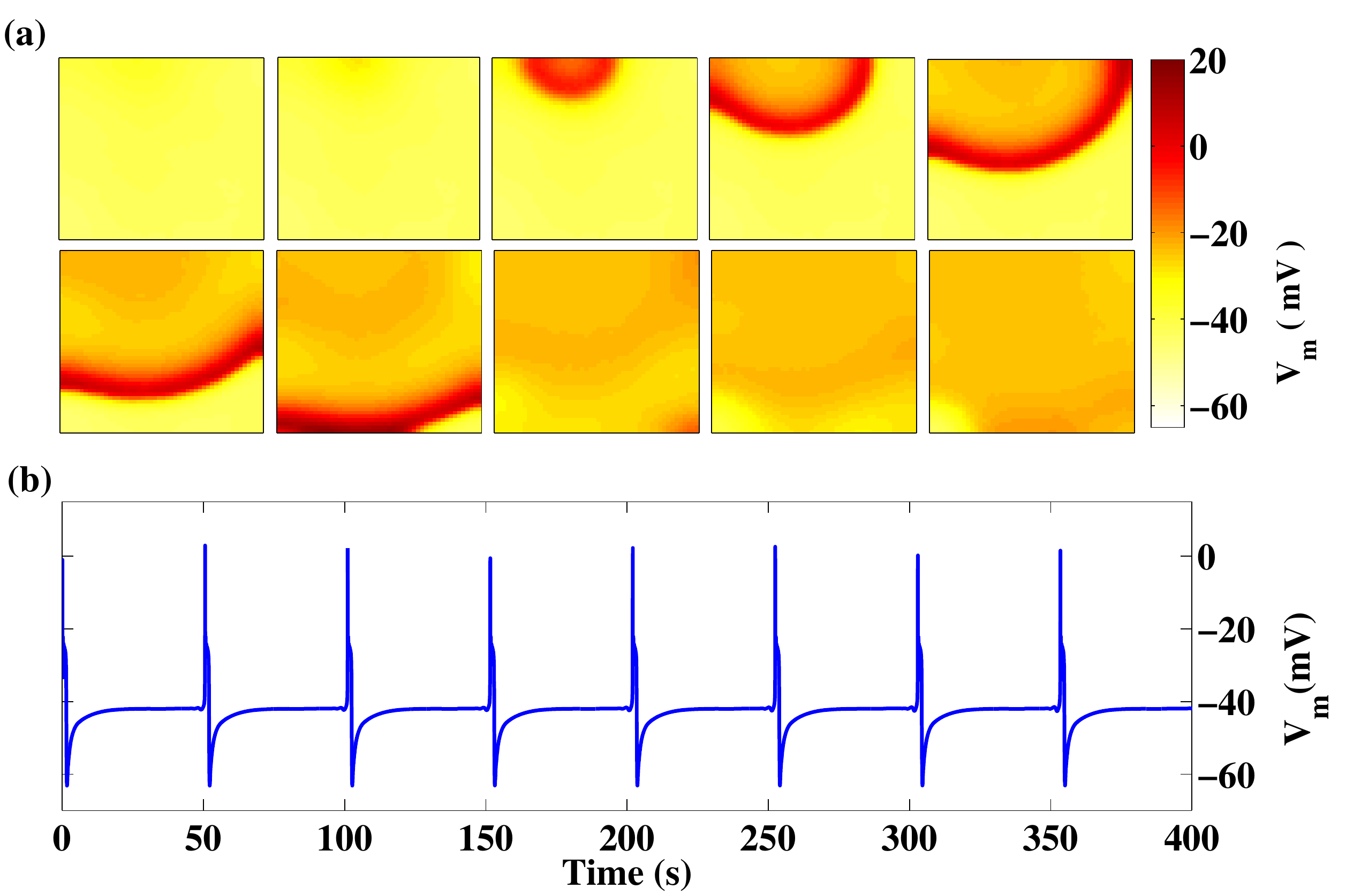} }
   \caption{ Regular periodic activity in the $2$-D lattice of
   coupled myocytes and passive cells, observed for an
   inter-myocyte coupling strength, 
   $G_m= 20nS$. (a) Waves are emitted periodically from a single,
   dominant region characterized by high passive cell density. 
   The snapshots are separated
   in time by $75$ ms. (b) This behaviour causes each cell in the
   system to exhibit a periodic pattern of activity with a period $T
   \sim 50$ s. The only difference between the recorded time series of
   any two cells in the system is a temporal shift, dependent on the
   proximity of the cell to the source.
}
   \label{fig:periodic_activity}
\end{figure}}

{\subsubsection{Emergence of irregular activity}
\label{subsub:irregular}}

Despite the fact that the activity of all oscillating cells in the LS
and GS regimes have the same effective time period, we note that this
does not imply that the activity of all cells are temporally
synchronized in these regimes, nor that they exhibit simple dynamical
behaviour. In fact, for values of the inter-myocyte gap junction
conductance in the range $6 nS\lesssim G_m \lesssim 19.2 nS$ the
region of parameter space characterized by GS occasionally
exhibits irregular activity: oscillations with a period of $\sim 15s$
that are erratically interrupted, for short durations, by oscillations
with a period of $\sim 1s$. This activity arises due to the
competition between strong and weak pacemaker-like regions that can,
on occasion, generate disordered activity in the form of transient
spiral waves. The strength and number of pacemaker-like regions
is strongly correlated to the passive cell distribution on the
lattice, and the possibility of competition between regions is
more significant for larger lattices~\cite{Xu2013}. 


%
{\begin{figure}[tb] 
   \centering
{\includegraphics[width=12.0cm]{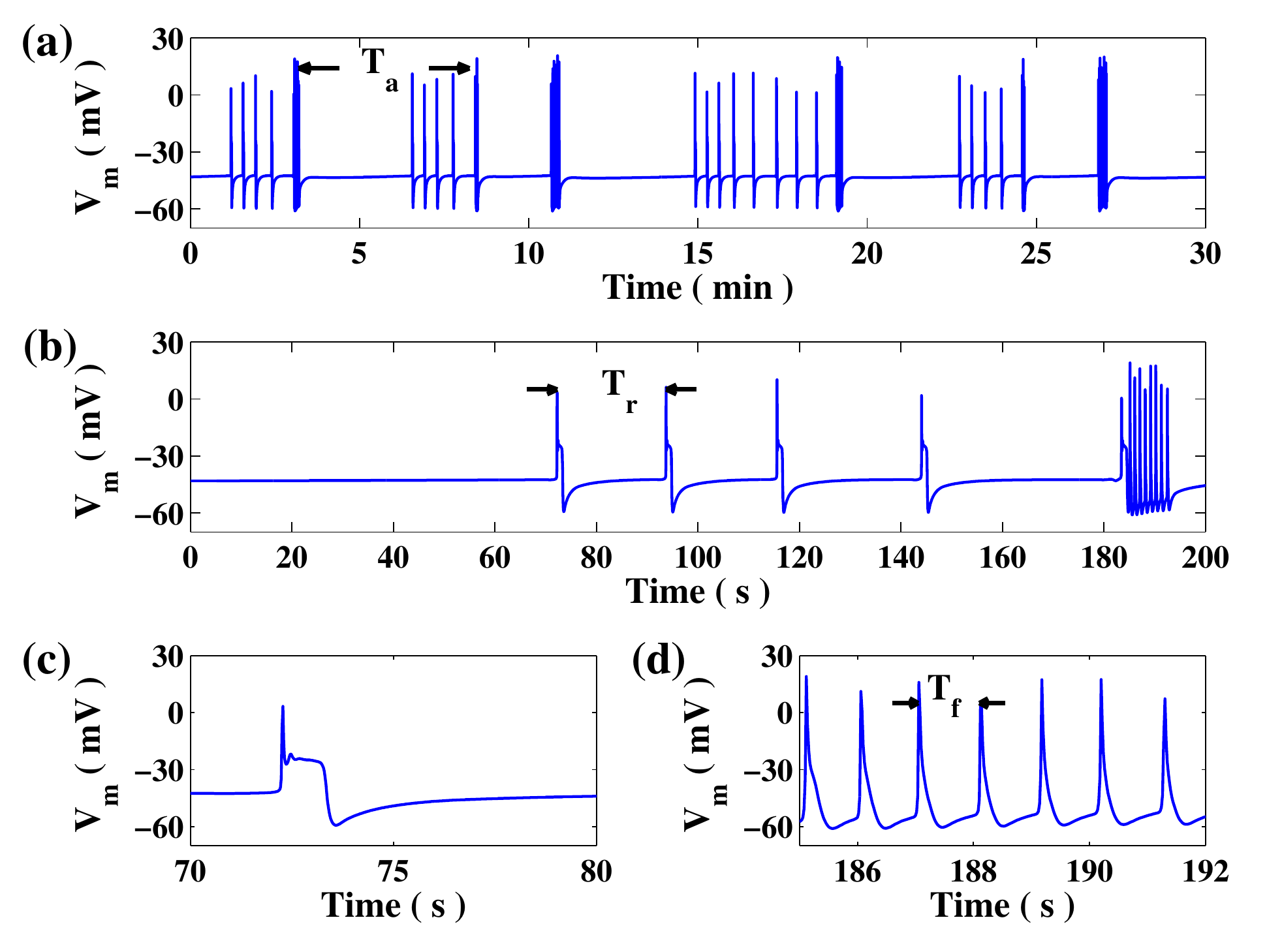}}
   \caption{Irregular patterns of activity in the $2$-D lattice
   of coupled myocytes and passive cells,
   observed for inter-myocyte coupling strength, $G_m = 12 nS$. 
   (a) The membrane potential of a single
   cell exhibits recurrent patterns of activity, each
   pattern arising after an interval $T_a$.
   (b) Each pattern is characterized by an
   initial quiescent phase,
   followed by a series of action potentials with
   period $T_r \sim 20 s$ and a brief duration of fast
   oscillations with periods $T_f \sim 1 s$. The profiles of a
   representative action potential (c) and fast oscillations (d) are
   also shown.
   }
   \label{fig:irregular}
\end{figure}}

As this irregular behaviour occurs in the GS regime, all cells
exhibit the same qualitative time series, with only a temporal shift.
Hence, we restrict our attention to the behaviour of a generic
cell in the lattice.  The evolution of the membrane potential of
a single randomly selected cell for $G_m = 12 nS$ 
is shown in Fig.~\ref{fig:irregular}.
We observe periodically occuring patterns of irregular activity, with
consecutive patterns separated by $T_a \sim 4$ min. The
structure of this irregular pattern is shown in
Fig.~\ref{fig:irregular}~(b), where a sequence of action potentials 
with a period $T_{r} \sim 20s$ are followed by a set of fast
oscillations of
period $T_f \sim 1s$. As seen from Fig.~\ref{fig:irregular}~(c-d), the
fast oscillations do not exhibit the ``plateau'' that characterizes action
potentials.
We find that the initial regular
activity (of period $T_r$) arises from waves generated by a single,
dominant ``pacemaker'' region which, as suggested by the detailed
analysis of a simplified model~\cite{Xu2013}, is characterized by a
high density of passive cells. 
%
{\begin{figure}[tb] 
   \centering
{\includegraphics[width=16.0cm]{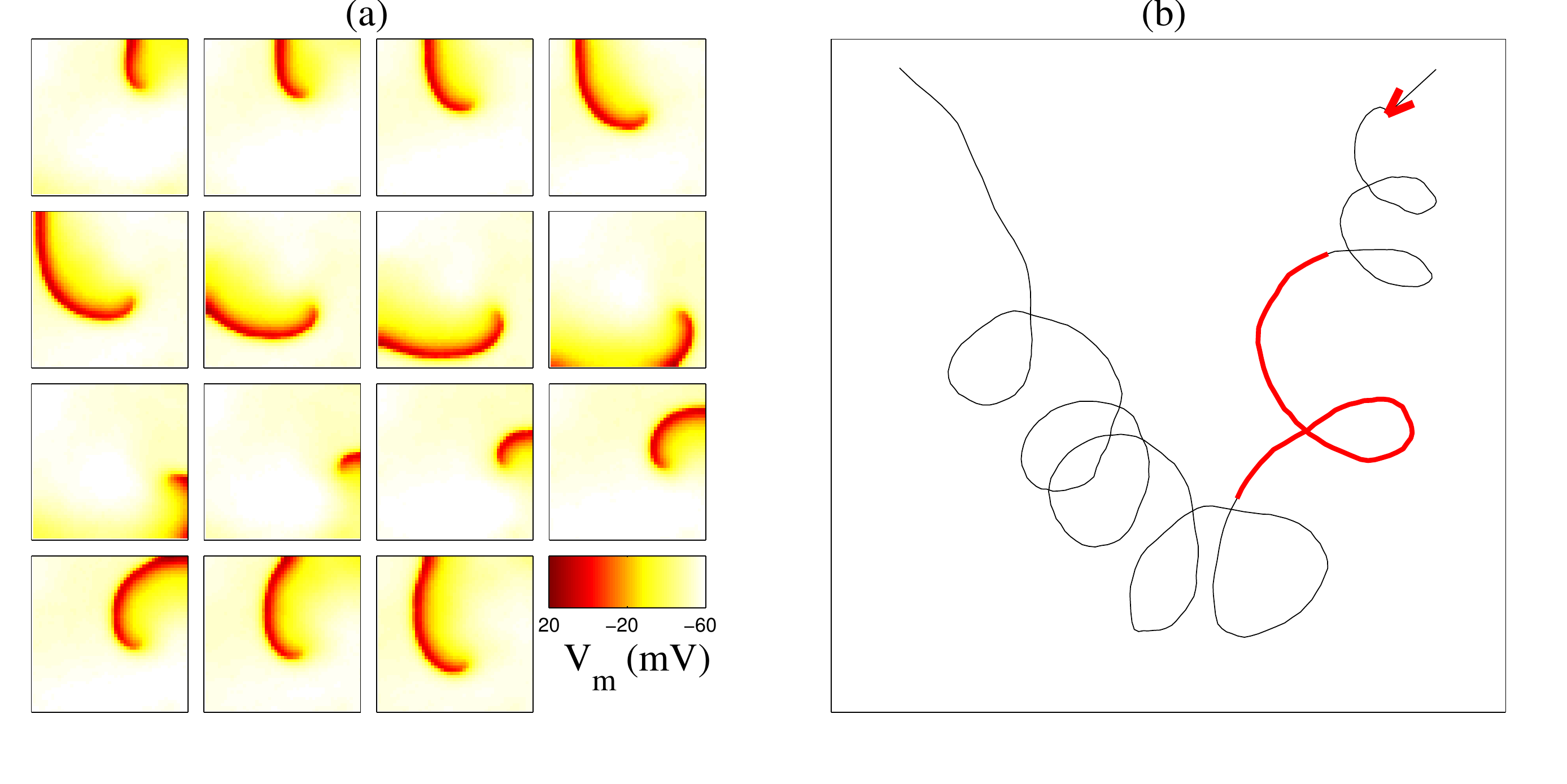}}
   \caption{ Occurrence of spiral waves of activity in a $50
   \times 50$ lattice of coupled myocytes and passive cells, observed
   for inter-myocyte coupling strength, $G_m=12nS$. (a) Snapshots of
   membrane potential are shown at intervals of $T = 100 ms$,
   in sequence from left to right and from top to bottom. 
   (b) Trajectory indicating the motion of the tip of the spiral
   on the 2-D lattice. The two arrows indicate the locations of
   the spiral when it emerges and disappears, respectively 
   The thick segment corresponds to the
   sequence shown in (a).
}
   \label{fig:irregular_spiral}
\end{figure}}

The irregular behaviour is a consequence of transient, recurrent
spiral wave activity (Fig.~\ref{fig:irregular_spiral}). The motion of
the spiral wave can be characterized by the trajectory of its tip,
which is defined here as the position on the lattice where the
membrane
potential is $V_m = -30 mV$, intermediate between resting and
depolarized states, and where the variable $h$, describing the
inactivation of sodium channels is equal to $h = 0.5$.
The motion of the spiral tip over a single rotation period
is shown in Fig.~\ref{fig:irregular_spiral}~b. 
We observe that the emergence of spiral activity is
not sensitive to the precise choice of model parameters.

\begin{figure}[tb] 
   \centering
{\includegraphics[width=16.0cm]{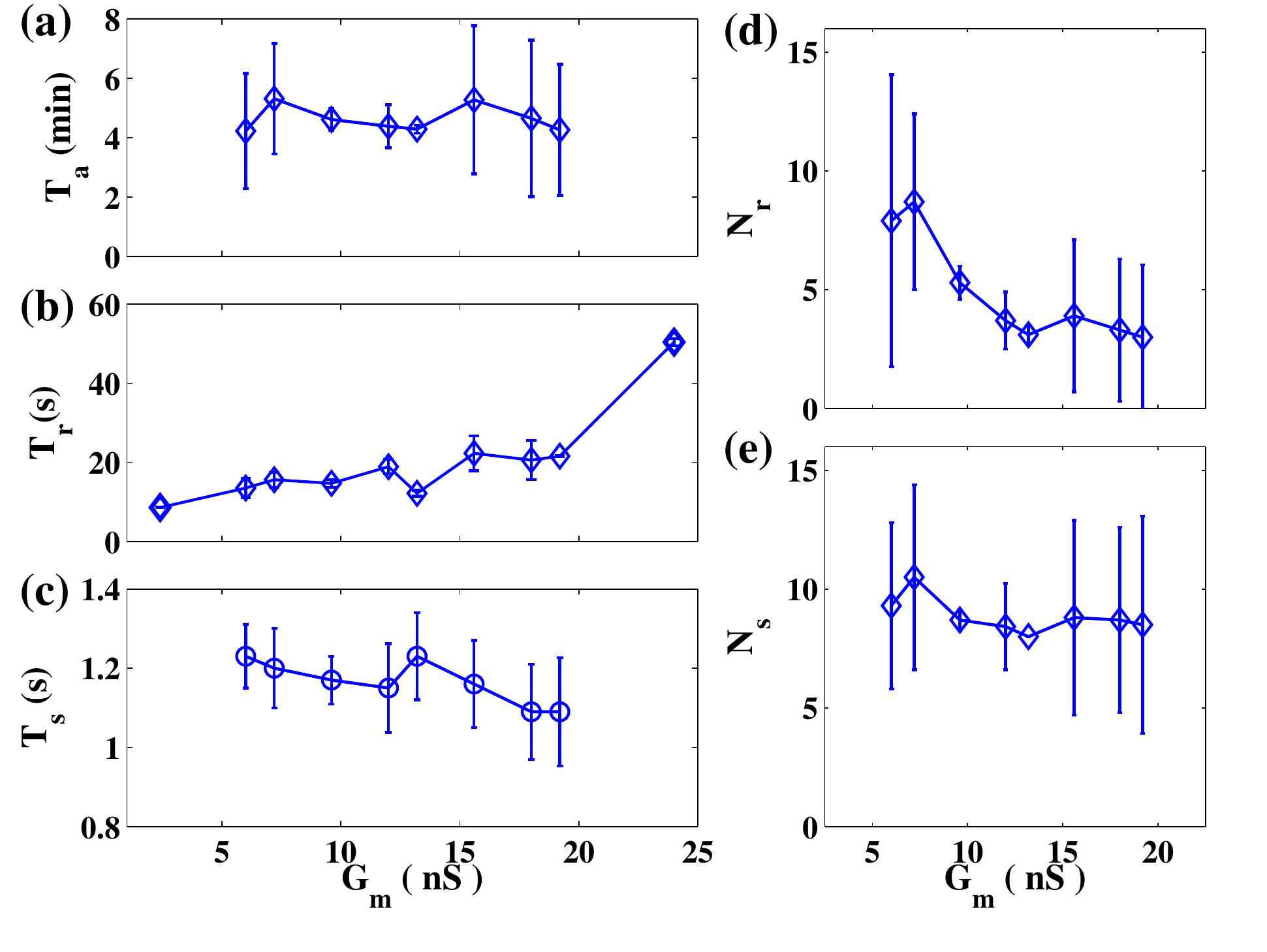}}
   \caption{ Dependence on $G_m$ of the quantities characterizing the
   patterns of activity shown in Fig.~\ref{fig:irregular}, namely the
   mean values of (a) $T_a$, (b) $T_r$, and (c), $T_f$ (c); and of the
   number of (d) action potentials, $N_r$, and (e) fast
   oscillations, $N_f$. The error bars indicate the standard deviation
   of the fluctuations of each individual quantity. Each quantity is
   measured over an interval of at least $6000$ s. {The
   quantities shown here have been determiend after averaging over 30
   independent realizations of the passive cell distribution. }
    }
   \label{fig:irregular_statistics}
\end{figure}

To quantitatively characterize the spiral dynamics, we identify
$T_f$ as the rotation period of the spiral and $T_a$ as the interval
between two successive appearances of transient spiral activity.
Additionally, $T_r$ is identified as the interval between successive
regular waves generated by the region with 
higher passive cell density. We
also define $N_r$ as the number of regular waves appearing
prior to the appearance of a spiral and $N_f$ as the
number of rotations by a spiral during its lifetime.
Fig.~\ref{fig:irregular} shows that the
values of the characteristic times $T_a$, $T_r$ and $T_f$, as well
as $N_r$ and
$N_f$, vary between successive irregular
patterns. Nevertheless, simulations over large time scales confirm that
the behaviour shown in Fig.~\ref{fig:irregular} is statistically
stationary. Fig.~\ref{fig:irregular_statistics} displays the
dependence of the mean of these quantities, obtained over
a sufficiently long time interval, on $G_m$, with the error bars
indicating the standard deviation. We observe that the mean
value of $T_a$ is approximately constant over the range $ 6 nS \le
G_m \le 19.2nS$. In contrast, the mean value of $T_r$ ($T_f$) slightly
increases (decreases). Despite the relatively large error bars for
$N_{r}$ and $N_{s}$, we find that these quantities remain more or less
constant, except for a decrease in $N_{r}$ at low $G_m$. 

%


\section{Discussion}


In this paper we address the question as to how spontaneous
contraction-inducing currents can emerge in the uterus despite the absence
of any clearly identifiable pacemaker cells in this organ.
Our proposed solution is motivated in part by the
experimental observation that
gap junction expression strongly increases towards the end of
pregnancy~\cite{Miller1989,Miyoshi1996}, leading to an increase in
inter-cellular coupling. We investigate the role of such coupling
in triggering spontaneous uterine activity by considering the 
coupling of excitable myocyte cells to electrically passive cells in a
mathematical model. We have used a realistic description of
myocyte activity based on a model recently developed by Tong \emph{et
al.}~\cite{Tong2011}, and have incorporated interaction between
myocytes and passive cells. We investigate the collective activity of
such an assembly both in the context of a single myocyte coupled to
one or more passive cells, as well as a 2-D lattice of
myocytes with nearest neighbor coupling with each myocyte interacting
with a randomly distributed number of passive cells.
While simplified models of excitable cells interacting with passive
cells have been considered in earlier investigations~\cite{Singh2012},
the use of physiologically realistic model in the present paper allows
us to make semi-quantitative predictions and {provides} a framework for
explaining uterine tissue activity both {\it in vitro} and {\it in
vivo}.

In addition to the
gap junctional expression, several other physiological parameters,
such as the sodium conductance, $g_{\rm Na}$, have
been observed to change significantly through the course of pregnancy~\cite{Yoshino1997}.
The experimentally observed values of
the resting potential of Interstitial Cajal-like cells (ICLC), as well
as of myocytes, are also found to vary over a wide range. For
the purposes of our simulations, we choose a set of parameters over a
range well within the bounds set by experimental observations.
We have confirmed that the results of our model simulations are
qualitatively robust to small variations in parameter values (see SI
for details). 

We find that the properties of a single myocyte coupled to a number of
passive cells are qualitatively very similar to the behaviour observed
using a simple excitable model description of
myocyte activity~\cite{Singh2012}. The coupling
between myocyte and passive cells, $G_p$ has been chosen to be of the order
of $\sim 1 nS$, which corresponds to the expression of $\sim 20$ gap
junctions of conductance $\sim 50~$ pS~\cite{Valiunas2004}. We note
that this is realistic, given the number of gap junctions known
to be expressed in a myocyte~\cite{Miyoshi1996}.


We observe that the quantitative nature of our numerical results
is dependent on the precise value of the passive cell resting
potential, $V_p^r$, as can be seen from Fig.~\ref{fig:osc_G_Vpr}.
It has been experimentally observed that the
resting potential of an ICLC is $-58 \pm 7$ mV.
As the fraction of ICLCs in the tissue does not exceed
$n_p \approx 0.2$ \cite{Popescu2007,Young2007}, the results
shown in Fig.~\ref{fig:osc_G_vs_np} suggest that no oscillations
would occur in a tissue containing only myocytes and ICLCs.
However, uterine tissue contains other electrically passive
cells, such as fibroblasts which have a much higher resting
potential, $V_p^F \approx -15 mV$, and are known to play an important
structural role in uterine tissue~\cite{Takemura2005,Malmstrom2007}.
A theoretical analysis, documented in the Supplementary Information,
suggests that a small population of such passive cells may result
in an ``effective" resting potential for the passive cells that is higher than that of
ICLCs alone. This provides a justification for the range of
values of $V_p^r$ used in the simulations reported here, which exhibit 
spontaneous oscillations in a system of coupled myocytes and passive
cells.


For the situation in which myocytes are coupled on a $2$-D lattice to
their nearest neighbors, as well as to a random number of passive
cells, we observe that a progressive increase in the coupling between
cells results in a gradual transition from quiescence to the
appearance of small clusters of oscillating cells. Further
increase in coupling causes these clusters to grow and merge until a
single cluster occupies the
entire system. These features are qualitatively
consistent with observations from experiments on a co-culture of
myocytes and fibroblasts~\cite{Chen2009}.
This transition to regular
global synchronization occurs via an interesting dynamical regime in
which transient, recurrent spiral waves propagate through the system
giving rise to activity with a period of $\sim 1s$. This regime of
irregular spatiotemporal behaviour has not been previously reported.
It is of interest to note that these patterns resemble the complex
waves seen in \textit{in vitro} experiments on guinea pig
uterine tissue performed by Lammers \emph{et al}~\cite{Lammers2008}.
Thus, although a precise quantitative comparison between simulations and
experiments would require a more exhaustive investigation, our
results are in close qualitative agreement with the
observed features of waves propagating in uterine tissue. 

We find that it takes approximately $340 ms$ for a wave to propagate
across a $2$-D myocyte assembly of size $50 \times 50$ cells.
Assuming that the length of a cell is $\sim 225 \mu
m$~\cite{Yoshino1997}, this corresponds to an action potential
propagation velocity of $\approx 3.3 cm/s$, which is consistent with
the values obtained in Ref.~\cite{Lammers1999,Lammers2008}. We
observe
that the period of fast activity is of the order of $\sim 1s$, which
is also in agreement with the results of Ref.~\cite{Lammers2008}.
Additionally, we note that the value of the inter-myocyte coupling
used to obtain the patterns shown in Figs.~\ref{fig:irregular} and
\ref{fig:irregular_spiral} is $G_m = 12 nS$, which corresponds to an
expression of $n_{gj} \approx 240$ gap junctions, a value
consistent with experimental observations~\cite{Miyoshi1996}.

%
For medical applications, the key question is to understand the
generation of force in the uterine tissue.
In myocytes, action
potentials induce a large influx of Calcium, which in turn leads to
cell contraction. Available models addressing the question of force
generation rest on bursts of Calcium influx inside
cells~\cite{Burdyga2007,Tong2011,Maggio2012}. 
It is an open question as to whether such activity is a result of
intrinsic electrophysiological dynamics of local cell clusters or due
to re-entrant waves propagating around the organ.
We note that in our simulation of a 2-D lattice of coupled myocytes
and passive cells, rapid spiking activity is
observed when the system exhibits spiral waves (Figs.~\ref{fig:irregular} and
\ref{fig:irregular_spiral}). Determining whether spirals induce
transient, pathological contractions, as is the case in the heart,
or are required to generate a strong force at the time of delivery,
cannot be answered without a better understanding, both at the
cellular and tissue level~\cite{Lammers2008,Lammers2013}.

The results reported in this paper present a picture that
is qualitatively, and to an extent quantitatively, consistent with a
number of experimental observations, despite the limitations
inherent to physiologically detailed models, such as the
uncertainties in the characterization of potentially crucial model
parameters. This suggests that the mechanism under
consideration, namely the electrical coupling between excitable
myocytes and passive cells, is at least partially responsible for the
generation of spontaneous electrical activity. Our work provides a
feasible and falsifiable hypothesis that suggests new avenues
for further investigation into this issue, such as the effect of
increasing sodium conductance, or the role of hormones
such as oxytoxin in the course of pregnancy.

\section*{Acknowledgement}
The work has been supported by the Indo-French Center for Applied Mathematics
program, and the JoRISS exchange program between the ENS-Lyon and the 
East China Normal University. 
JX is partly supported by NFSC under grand 11405145.
SNM is supported by the IMSc Complex Systems Project.
AP is grateful to the Humboldt foundation for support.
We thank IMSc
for providing access to the ``Annapurna'' supercomputer.


\section*{Author contributions}
Conceived and designed the numerical work: JX, SNM, RS, NBG, SS and AP.
Performed the numerical work: JX and SNM.
Analyzed the data: JX, SNM, NBG, SS and AP.
Contributed to analysis tools: RS and NBG.
Wrote the paper: JX, SNM, NBG, SS and AP.

\bibliography{uterine_elec}

\setcounter {section} {0}
\setcounter {subsection} {0}
\setcounter {table} {0}
\setcounter {figure} {0}
\setcounter {equation} {0}

\renewcommand{\thesection}{S\arabic{section}} 
\renewcommand{\thetable}{S\arabic{table}}
\renewcommand{\thefigure}{S\arabic{figure}}
\renewcommand{\theequation}{S\arabic{equation}}


\newpage

\begin{center}
{\Huge{SUPPORTING INFORMATION}}
\end{center}
\vspace{0.2cm}

\section{Introduction}
\label{sec:S1}

This document contains supporting information for the manuscript ``The role of cellular coupling in the spontaneous generation of electrical activity in uterine tissue'', and is organized as follows. In section~\ref{sec:S2}, we discuss a model for the electrical activity of uterine myocyte cells, developed by Tong \etal~\cite{Tong2011}, that forms the basis for the model used in the manuscript. In particular, we review their description of the Ca$^{2+}$ dynamics and highlight a set of discrepancies between their model, as described in the supporting information (SI) of \cite{Tong2011}, and the source code accompanying their manuscript. We subsequently describe the precise form of the model used for our simulations. Next, in section~\ref{sec:S3}, we demonstrate that the model used here exhibits the expected response to an external stimulus and qualitatively reproduces results from the electrophysiological voltage clamp experiments referred to in \cite{Tong2011}. Finally, section~\ref{sec:S4} is devoted to the study of the combined effect of the different types of passive cells that are known to be present in uterine tissue.


\section{Explicit description of the current associated with the Na$^+$-Ca$^{2+}$ exchanger}
\label{sec:S2}

The equation governing the intracellular calcium concentration, $\Ca$, in the model of uterine smooth muscle cells developed by Tong \etal~\cite{Tong2011} is:
\begin{equation}\label{equ:1}
\frac{\dd\Ca}{\dd t} = -(J_{\rm{Ca, mem}} + J_{\rm{NaCa}} + J_{\rm{PMCA}})\,,
\end{equation}
where $J_{\rm{Ca, mem}}$ includes all the Ca$^{2+}$ currents from the channels that are expressed at the membrane; $J_{\rm{NaCa}}$ is the flux from the Na$^+$-Ca$^{2+}$ exchanger, and $J_{\rm{PMCA}}$ is the flux related to the process of plasmalemmal Ca$^{2+}$-ATPase. 
The Na$^+$-Ca$^{2+}$ exchanger contributes to approximately $30\%$ of the extrusion of $\Ca$ \cite{Tong2011}. It carries one Ca$^{2+}$ out of, and brings three Na$^+$ in to, the cell. Thus, the role of the exchanger is: (1)~to reduce $\Ca$, and (2)~to induce a net inward current that depolarizes the membrane potential. Using the standard convention, namely that outward ionic currents are taken as positive~\cite{Junge1992}, the influx of Ca$^{2+}$ into the cytosol results in a negative electric current. We note that there exists a discrepancy between the sign of $J_{\rm NaCa}$ in the SI of \cite{Tong2011} and in the source code accompanying that manuscript, and this issue has been attended to in our simulations. Furthermore, as discussed in detail in Sec.~\ref{sub:S3.2}, the description of the Na$^+$-Ca$^{2+}$ exchanger in the source code of~\cite{Tong2011} corresponds to Ca$^{2+}$ influx into the cell, while that presented in the SI of~\cite{Tong2011} corresponds to the extrusion of Ca$^{2+}$. We implement the latter in our simulations, as it is physiologically correct.

Finally, as seen in Table~\ref{tab:Na-Ca}, there exists a slight discrepancy between the values of K$_{m,\rm Allo}$ (denoted as $K_{mCaAct}$ in~\cite{Weber2001}), displayed in the SI of \cite{Tong2011} and in the source code accompanying that manuscript. In our simulations, we use a value of K$_{m, \rm Allo}$ that lies between the values displayed in the SI of \cite{Tong2011} and the value used in Weber \etal~\cite{Weber2001}, whose description of the Na$^+$-Ca$^{2+}$ exchanger was the basis of that used by Tong {et al.}~\cite{Tong2011}. In addition, the values of $\bar{J}_{NaCa}$, $K_{m,\rm Nai}$, $K_{m,\rm Cai}$ and $n_{Allo}$ are taken from the corresponding values of $V_{max}$, $K_{mNai}$, $K_{mCai}$ and $n_{Hill}$ in~\cite{Weber2001}. We note, however, that the results obtained using these values are qualitatively and quantitatively similar to those obtained using the values in the code of Tong \etal~\cite{Tong2011}.
\begin{table}[h!]
\centering
\begin{tabular}{|c|c|c|c|c|}
\hline
Variables                    & Tong \etal~\cite{Tong2011} (SI) & Tong \etal~\cite{Tong2011} (code) & Weber \etal~\cite{Weber2001} & Present work \\        \hline
$\bar{J}_{NaCa}$(pA/pF) & $11.67$                         & $11.67$                           & $22.6$                       & $22.6$ \\              \hline
K$_{m,Allo}$(mM)             & $3\times10^{-3}$                & $3\times10^{-4}$                  & $1.25\times10^{-4}$          & $1.25\times10^{-3}$ \\ \hline
K$_{m,\rm Nai}$(mM)          & $30$                            & $30$                              & $12.3$                       & $12.3$ \\              \hline
K$_{m,\rm Cai}$(mM)          & $0.007$                         & $0.007$                           & $0.0036$                     & $0.0036$ \\            \hline
$n_{Allo}$                   & $4$                             & $4$                               & $2$                          & $2$ \\                 \hline
\end{tabular}
\caption{
Values of the parameter used in the description of the Na$^+$-Ca$^{2+}$ exchanger. The description of the Na$^{+}$-Ca$^{2+}$ exchanger in Tong \etal~\cite{Tong2011} was based on the approach of Weber \etal~\cite{Weber2001}, and we use the latter values in the current work, with the exception of K$_{m,Allo}$, whose value was chosen to lie between the corresponding values used in Weber \etal~\cite{Weber2001} and that displayed in the SI of Tong \etal~\cite{Tong2011}.
}
\label{tab:Na-Ca}
\end{table}


\section{Validation of the model}
\label{sec:S3}

\subsection{Voltage clamp experiments}
\label{sub:S3.1}

\begin{figure}[H]
\centering
\subfigure[]{\includegraphics[width=0.33\textwidth]{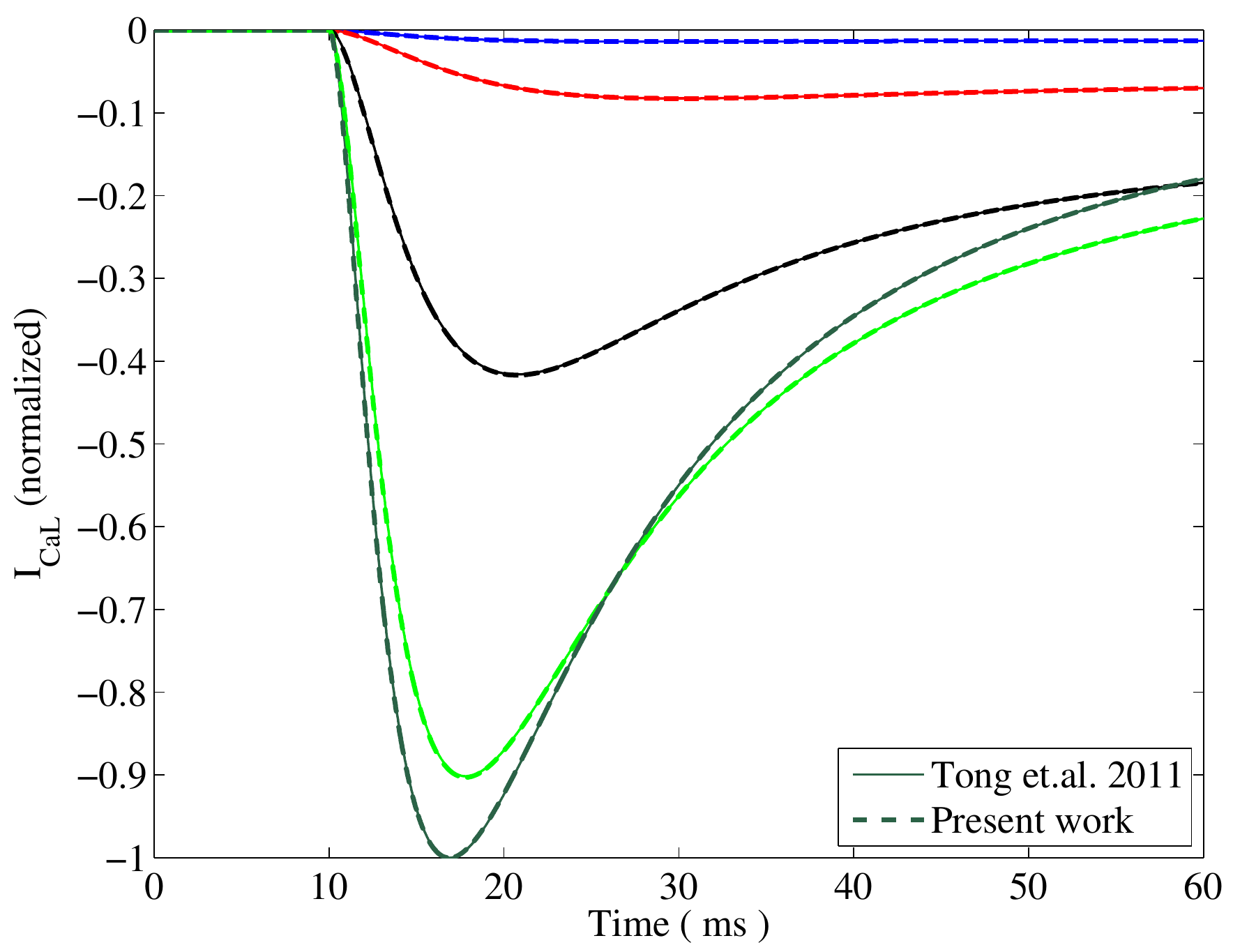}}
\subfigure[]{\includegraphics[width=0.33\textwidth]{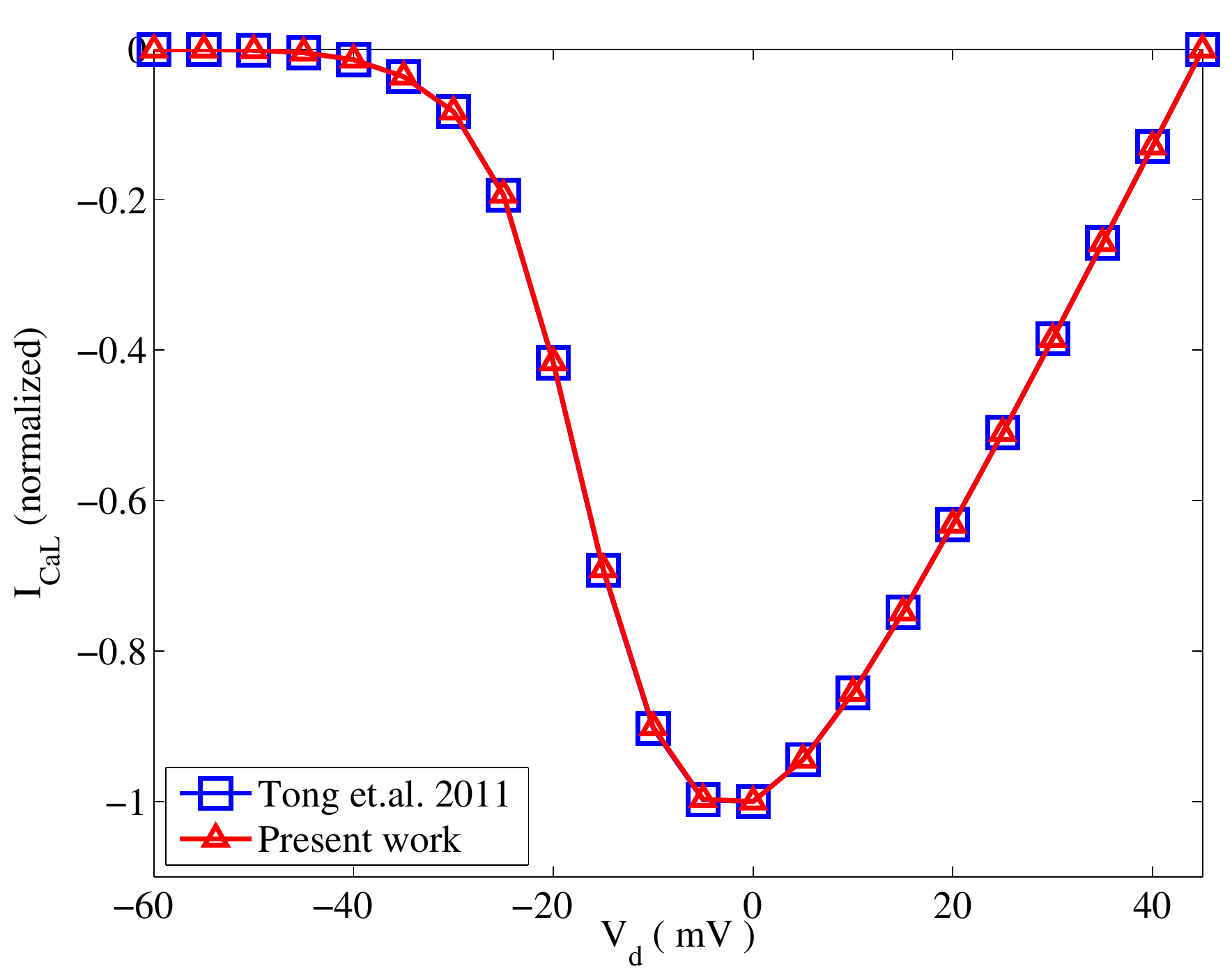}}
\subfigure[]{\includegraphics[width=0.33\textwidth]{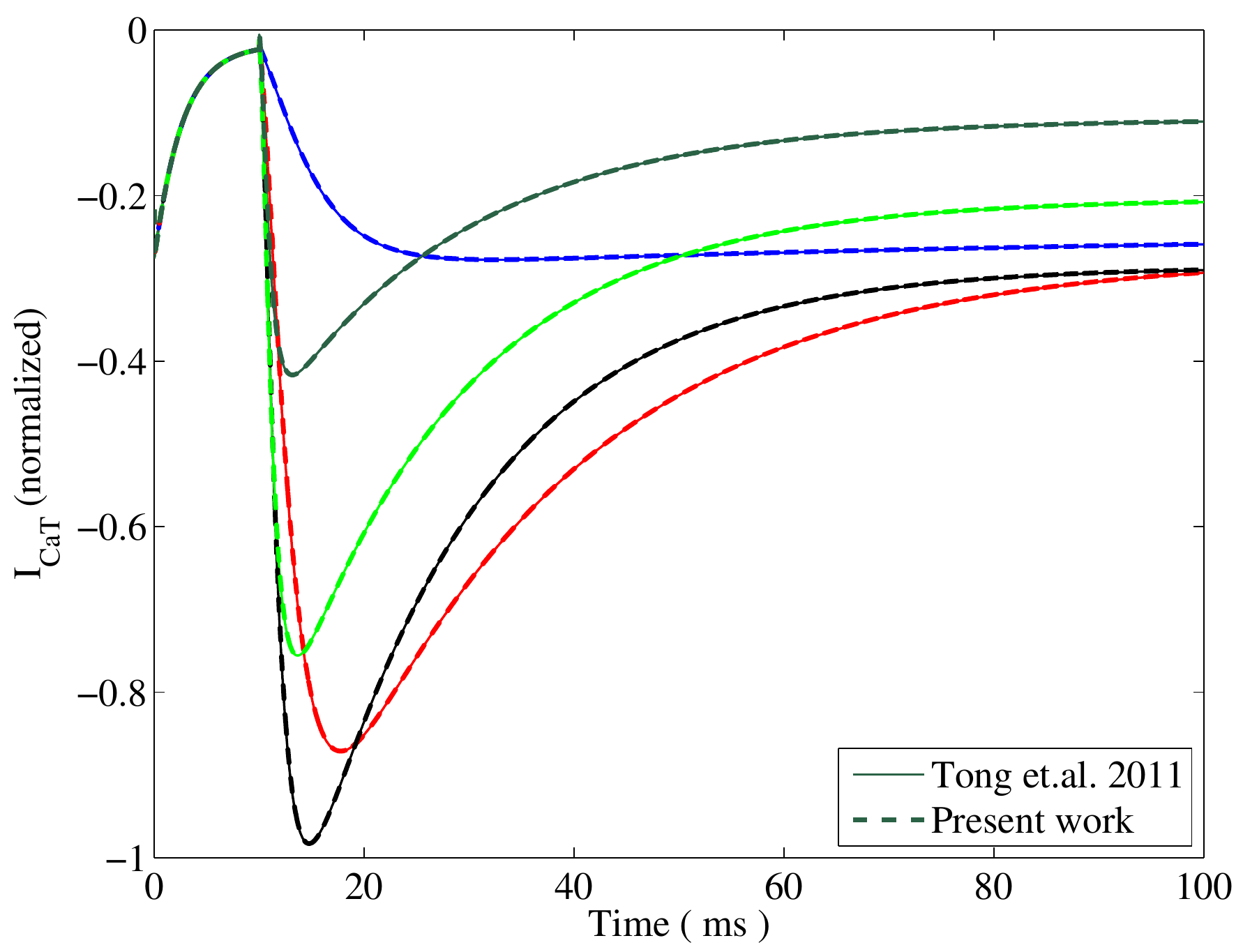}}
\caption{
Simulated voltage-clamp experiments on Ca$^{2+}$ channels. (a)-(b) Behaviour of L-type Ca$^{2+}$ channel current, $I_{CaL}$, for different depolarizing potentials in the range $-40$~mV to $0$~mV at voltage steps of $10$~mV with a holding potential $V_h=-60$mV, shown both as a function of (a) time and (b) depolarizing potential $V_d$, superimposed with results obtained using the model of Tong \etal~\cite{Tong2011}. (c) Behaviour of T-type Ca$^{2+}$ channel current, $I_{CaT}$, for different depolarizing potentials in the range $-60$~mV to $20$~mV, with a holding potential $V_h=-80$mV.
}
\label{fig:fig_voltageclamp}
\end{figure}

\begin{figure}[h!]
\centering
\subfigure[]{\includegraphics[width=0.33\textwidth]{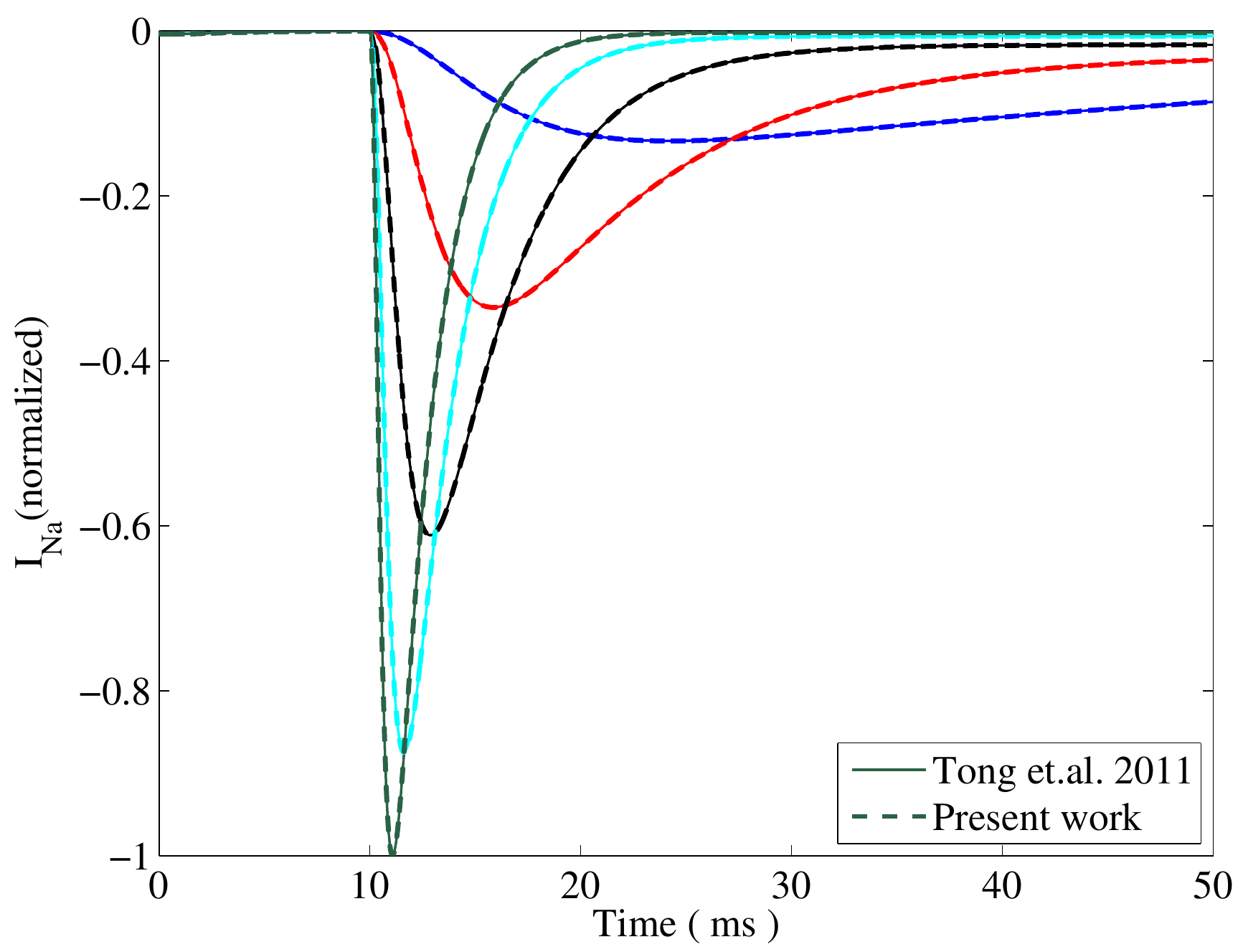}}
\subfigure[]{\includegraphics[width=0.33\textwidth]{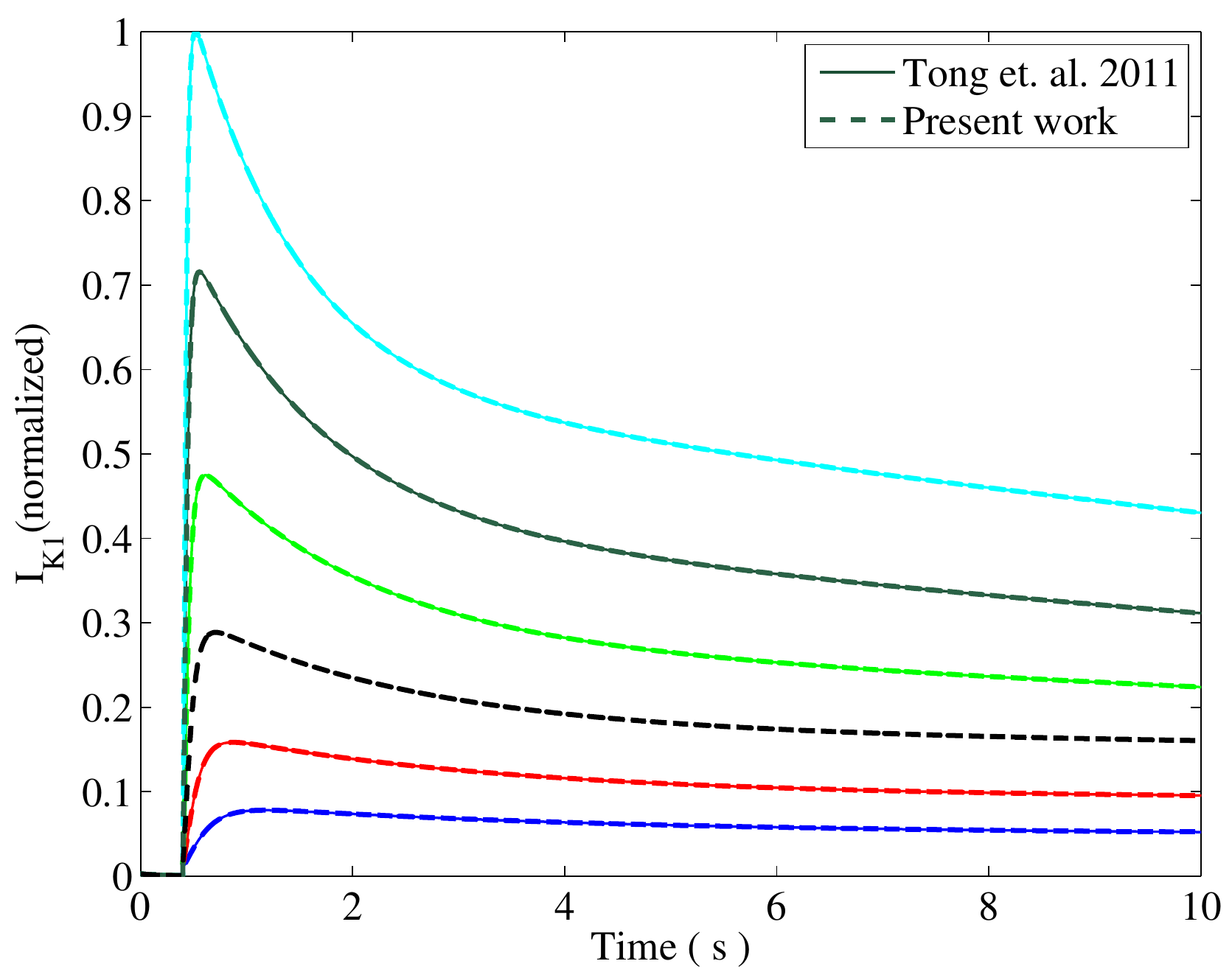}}
\subfigure[]{\includegraphics[width=0.33\textwidth]{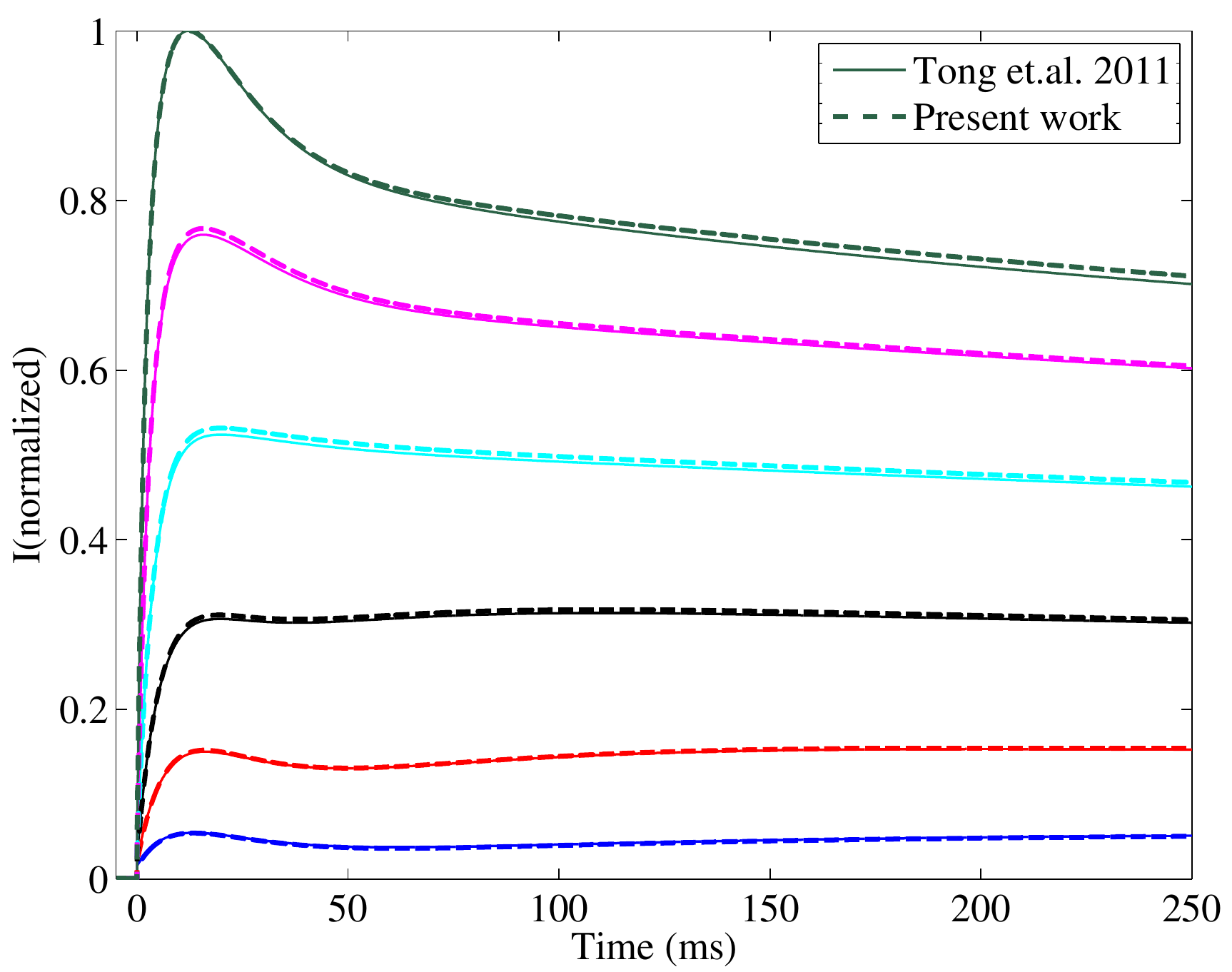}}
\caption{
Simulated voltage-clamp experiments on Na$^+$ and K$^+$ channels. (a) Behaviour of Na$^+$ channel current, $I_{Na}$, at different depolarizing potentials in the range $-40$~mV to $20$~mV with a holding potential $V_h=-40$~mV. (b) Behaviour of the K$^+$ channel current $I_{K1}$ for $g_k=0.8nS/pF$, at different depolarizing potentials in the range $-40$~mV to $10$~mV with a holding potential $V_h=-80$~mV, normalized to the peak current at $10$~mV. (c) Behaviour of the total K$^+$ channel current, at different depolarizing potentials in the range $-30$~mV to $70$~mV with a holding potential $V_h=-80$~mV, normalized to the peak current at $V_{d}=70$~mV.
}
\label{fig:fig_voltageclamp2}
\end{figure}

We validate our model by numerically simulating a voltage clamp experiment on the L-type Ca$^{2+}$ channel using a holding potential $V_h=-60$~mV. The recorded variations of $I_{\rm{CaL}}$ at  different depolarizing potentials (from $-40$~mV to $0$~mV with a step of $10$~mV) are shown in Figure~\ref{fig:fig_voltageclamp}~(a)-(b). It can be seen that the model reproduces the same current trace as the model of Tong \etal~\cite{Tong2011}. Similarly, Figure~\ref{fig:fig_voltageclamp}~(c) reveals that simulations of current traces recorded from voltage clamp experiments on T-type Ca$^{2+}$ channel reproduce the results of Tong \etal~\cite{Tong2011}. Furthermore, as seen in Figure~\ref{fig:fig_voltageclamp2}, numerical simulations of voltage clamp experiments on Na$^{+}$ and K$^{+}$ channels using our model reproduce the corresponding results obtained with the model of Tong \etal~\cite{Tong2011}.


\subsection{Current clamp experiments}
\label{sub:S3.2}

\begin{figure}[h!]
\centering
$\begin{array}{c}
\subfigure[]{\includegraphics[width=0.7\textwidth]{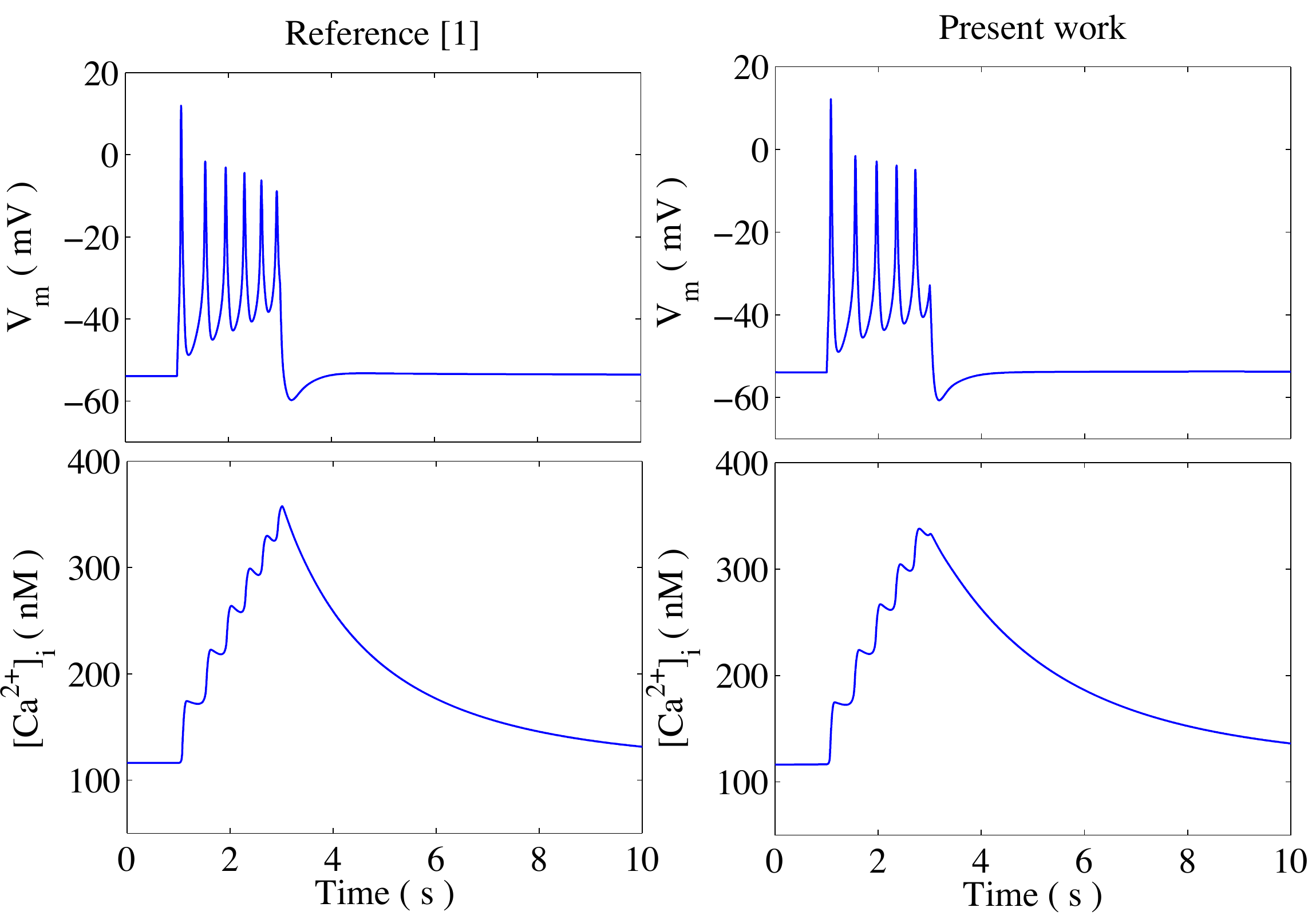}}\\
\subfigure[]{\includegraphics[width=0.7\textwidth]{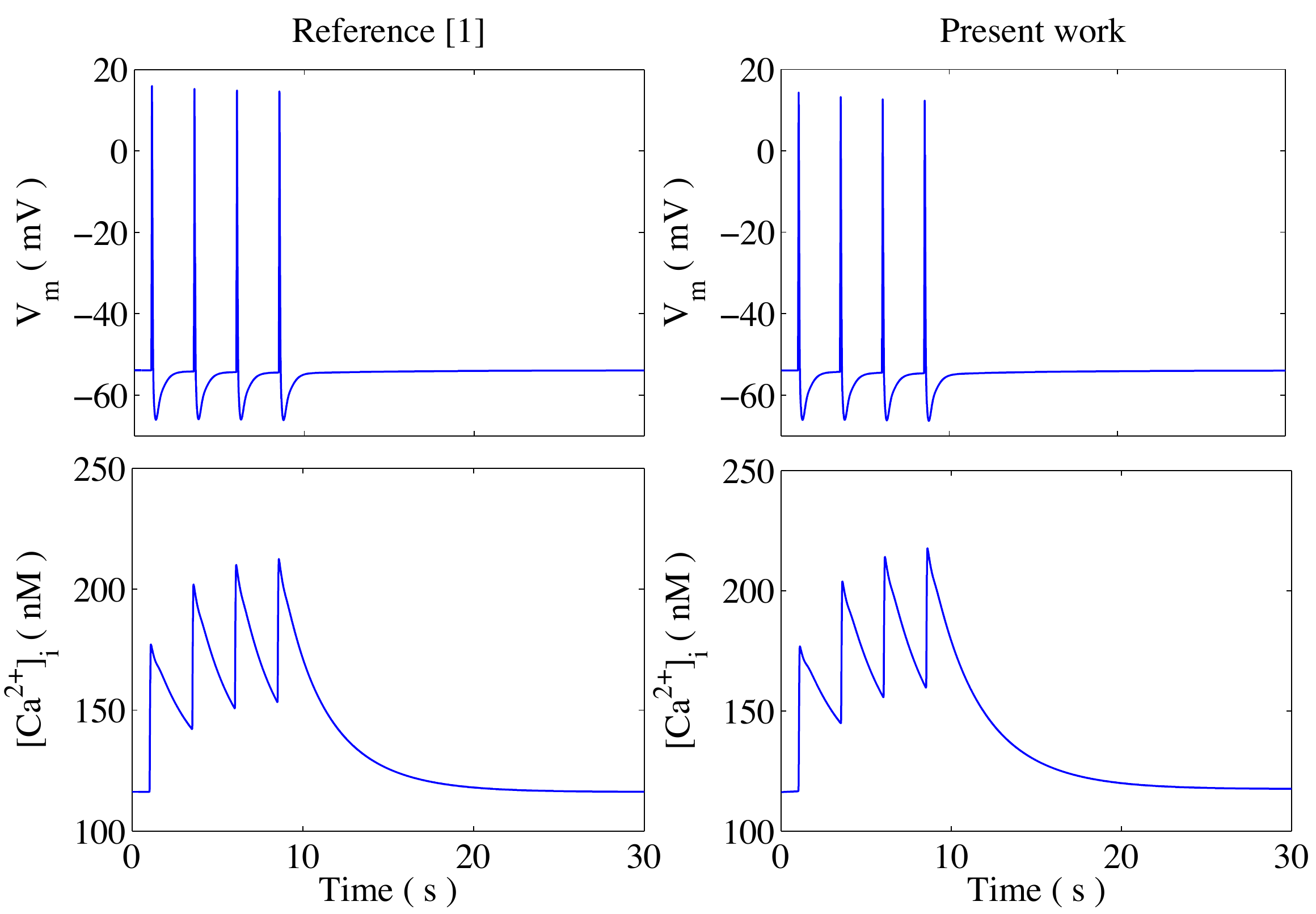}}
\end{array}$
\caption{
Action potentials obtained using our model, compared with the corresponding results obtained using the model of Tong \etal~\cite{Tong2011}, for the situations where: (a) A depolarizing current clamp of amplitude $I_{st}=-0.5pA/pF$ is applied for two seconds under control conditions (c.f. Figure 12 of Tong \etal~\cite{Tong2011}). (b) A stimulus of amplitude $-1.5$~pA/pF is applied over $20$~ms at $0.4$~Hz (c.f. Figure 13 of Tong \etal~\cite{Tong2011}).
}
\label{fig:current_clamp}
\end{figure}

In order to verify the response of our model to an external stimulus, we apply current pulses of the type used in~\cite{Tong2011}. As seen in Figure~\ref{fig:current_clamp}~(a)-(b), the activity of the membrane potential observed when simulating the current model is qualitatively similar to that obtained using the model of Tong \etal~\cite{Tong2011}.
It was reported in Tong \emph{et al.}~\cite{Tong2011} that a constant stimulus applied to the myocyte over a short duration only gives rise to oscillations when the conductance of the fast sodium channel, $g_{Na}$, is extremely low. Indeed, as seen in Fig.~\ref{fig:osc_region_{I}st_gk}, when a constant stimulus is applied to the model of \cite{Tong2011} at large values of the sodium channel conductance, a very large depolarization of the membrane potential, and consequently an excessively large value of the Ca$^{2+}$ concentration inside the cell ($\sim 610 mM$) is observed. When the stimulus current is subsequently turned off, one does not observe a return to the resting values of the membrane potential or the internal Ca$^{2+}$ concentration. This behaviour is potentially problematic, as the sodium channel conductance, which is very low in most smooth muscle cells, has been observed to increase significantly in rat myometrium during gestation~\cite{Ohya1989,Sperelakis1992}. Moreover, we observe that the model of \cite{Tong2011} does not exhibit spontaneous oscillations, even for large stimuli.
\begin{figure}[t]
\centering
{\includegraphics[width=\textwidth]{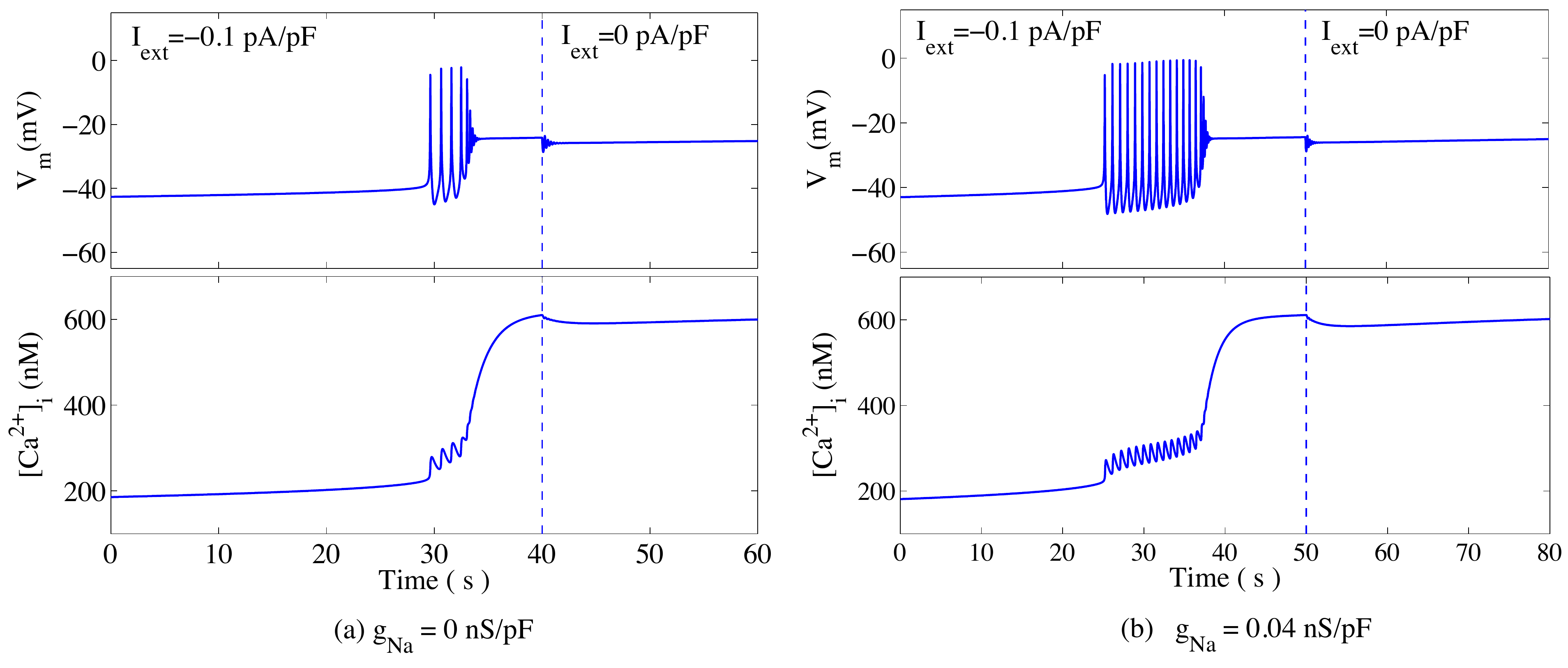}}
\caption{
Behaviour of the model of Tong \etal~\cite{Tong2011}, for the situation where a constant stimulus ($I_{st}=-0.1$~pA/pF) is applied for two values of the sodium conductance, viz. (a) $g_{Na} = 0$~nS/pF, and (b) $g_{Na} = 0.04$~nS/pF. The evolution of the [top] membrane potential, and [bottom] intracellular calcium concentration is displayed in each case. The vertical dashed line indicates the time at which the stimulus is turned off.
}
\label{fig:osc_region_{I}st_gk}
\end{figure}
In this regard, it can be observed that the description of the Na$^{+}$-Ca$^{2+}$ exchanger used in this work will, on the other hand, permit sustained oscillations that can be maintained as long as the external current is applied, without any drift of the Ca$^{2+}$ concentration (see Fig.~\ref{fig:oscillationg_modified_model}). Moreover, when the stimulus current is stopped, the myocyte membrane returns to its steady state. This observed activity provides justification for the form of the $I_{tot}$ equation discussed in Sec.~\ref{sec:S2}.
\begin{figure}[t]
\centering
{\includegraphics[width = 7.5cm]{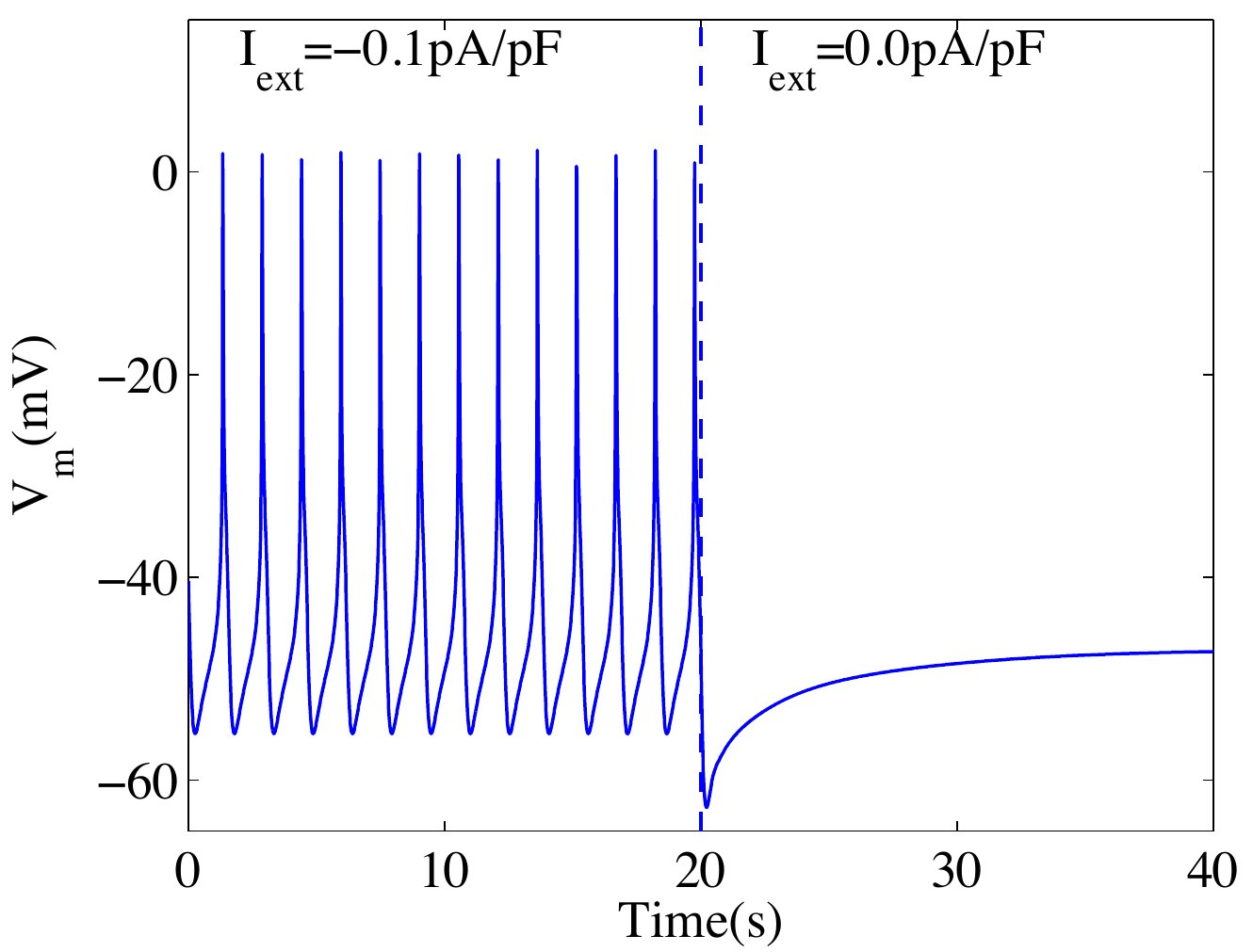}}
\caption{
Behaviour of our model for the situation where a constant stimulus ($I_{st} $=-0.1pA/pF) is applied. When the current is turned off (at the time indicated by vertical dashed line), the oscillations cease, and the system eventually returns to its resting state.
}
\label{fig:oscillationg_modified_model}
\end{figure}


\section{Description of ``effective'' passive cell dynamics}
\label{sec:S4}

The evolution of the membrane potential, $V_{g}$, of a generic passive cell, $g$, is captured by the equation:
\begin{equation}
- C_{g}\frac{\dd V_{g}}{\dd t } = G^{\rm int}_{g}\,(V_{g}-V_{g}^{r})\,,
\end{equation}
where $C_{g}, G^{\rm int}_{g}$ and $V_{g}^{r}$ are the membrane capacitance, conductance and resting potential of the passive cell, respectively. It is known that in addition to ICLCs, uterine tissue is populated by various other passive cells such as fibroblasts~\cite{Popescu2007}, although their numbers are much fewer than ICLCs \cite{Popescu2007}. The activity of uterine myocytes, which are typically coupled to ICLCs and fibroblasts through gap junctions~\cite{Kohl2005, Chilton2007, Duquette2005}, can hence be expressed as:
\begin{subequations}
\label{equ:couple}
\begin{align}
-C_{m}\frac{\dd V_{m}}{\dd t} = & I_{\rm ion} + n_{I}\,G_{p}\,(V_{m}-V_{I}) + n_{F}\,G_{p}\,(V_{m}-V_{F})\,,\label{eq:myocyte}\\
-C_{I}\frac{\dd V_{I}}{\dd t} = & G^{\rm int}_{I}\,{(V_{I}-V_{I}^{r})} + G_{p}\,(V_{I}-V_{m})\,,\label{eq:ICLC}\\
-C_{F}\frac{\dd V_{F}}{\dd t} = & G^{\rm int}_{F}\,{(V_{F}-V_{F}^{r})} + G_{p}\,(V_{F}-V_{m})\,,\label{eq:fibro}
\end{align}
\end{subequations}
where $C_{I}$ ($C_{F}$), $G^{\rm int}_{I}$ ($G^{\rm int}_{F}$) and $V_{p}^I$ ($V_{p}^{r}$) denote the capacitances, conductances and resting membrane potentials of ICLCs (fibroblasts) and where each myocyte is, on average, attached to $n_{I}$ ($n_{F}$) such cells. For simplicity, we assume here that the coupling strength between the excitable and passive cells, $G_{p}$, is independent of the passive cell type. As we describe below, the effective passive cell dynamics in the tissue can be modelled by considering the activity of a hypothetical ``combined'' passive cell.


\subsection{Effective passive cell}
\label{sub:S4.1}

It is known that myocytes (size $\sim 210 \mu$m \cite{Yoshino1997}), are typically much larger than ICLCs (size $\sim 75\mu$m, \cite{Popescu2007}) or fibroblasts (size $\sim 8\mu$m, \cite{Shibukawa2005}). Hence, the membrane capacitance of myocytes ($C_{m} \sim 120\mu$m \cite{Duquette2005}) is much larger than that of passive cells ($C_{I} \sim 85\mu$m \cite{Duquette2005} and $C_{F}\sim 4.5\mu$m \cite{Shibukawa2005}). As the relaxation time of a cell is proportional to its capacitance, it is to be expected that the membrane potentials of the aforementioned passive cells evolve much faster than that of myocytes. As this expectation is crucial in obtaining an effective description of the combined effect of several kinds of passive cell types, we first demonstrate its validity.

To this end, we compute the relaxation time in a system consisting of a myocyte, coupled to $n_{F}$ fibroblasts as well as to $n_{I}$ ICLCs. We began by considering the case where the deviation from the resting membrane potential of the myocytes ($V^{r}_{m}$) is small. In this situation, the ionic current, $I_{\rm ion}$, in the full equation (\ref{eq:myocyte}) can be linearized:
\begin{align}\label{equ:mycote_around_rest}
- C_{m} \frac{\dd V_{m}}{\dd t} =    G^{\rm int}_{m}\,(V_{m} - V_{m}^{r})  +
n_{I}\, G_{p}\,(V_{m} - V_{I}) +
n_{F}\,G_{p}\,\,(V_{m} - V_{F})\,,
\end{align}
and, together with Eq. \ref{equ:couple}(b,c), the equations for the membrane potentials can hence be written as:
\begin{equation}
\frac{\dd}{\dd t}\begin{pmatrix}
V_{m}\\
V_{I}\\
V_{F}
\end{pmatrix}  =  {\mathbf{M}}  \cdot
\begin{pmatrix}
V_{m} \\
V_{I}\\
V_{F}
\end{pmatrix}
+
\begin{pmatrix}
\frac{G^{\rm int}_{m}}{C_{m}}V_{m}^{r} \\
\frac{G^{\rm int}_{I}}{C_{I}}V_{I}^{r}\\
\frac{G^{\rm int}_{F}}{C_{F}}V_{F}^{r}
\end{pmatrix}\,,
\label{eq:matrix_equation} 
\end{equation}
where the matrix $\mathbf{M}$ describing the relaxation of the system is given by:
\begin{equation}
\mathbf{M}  =
\begin{pmatrix}
-\frac{1}{C_{m}}(G^{\rm int}_{m}+n_{I}G_{p}+n_{F}G_{p}) & \frac{n_{I}G_{p}}{C_{m}}&\frac{n_{F}G_{p}}{C_{m}}\\
\frac{G_{p}}{C_{I}} &-\frac{G^{\rm int}_{I}+G_{p}}{C_{I}}& 0\\
\frac{G_{p}}{C_{F}} & 0 &-\frac{G^{\rm int}_{F}+G_{p}}{C_{F}}
   \end{pmatrix}\,.
   \label{eq:def_mat_M}
\end{equation}
The solution of Eq.\eqref{eq:matrix_equation} can be written as a sum of three exponentially decaying solutions, of the form $\exp( - \lambda_i t)$, $i = 1, 2 , 3$. The decay rates, $\lambda_i$ are obtained by diagonalizing the $3 \times 3$ matrix $\mathbf{M}$, defined by Eq.~\ref{eq:def_mat_M}. As the matrix $\mathbf{M}$ is not symmetric, we note that its eigenvectors are not orthogonal to each other.

\begin{figure}[h]
\centering
\includegraphics[width=0.5\textwidth]{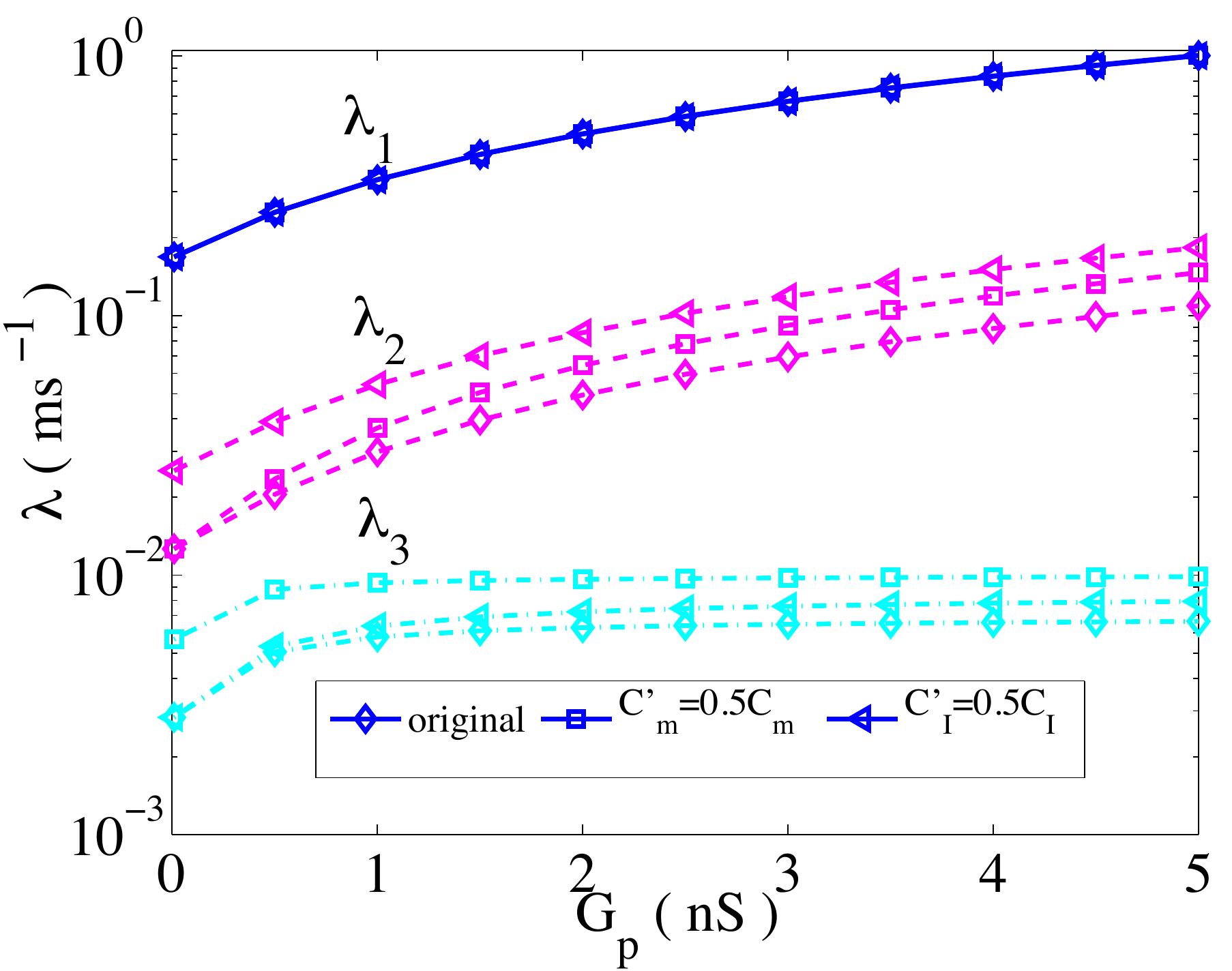} 
\caption{Decay rates $\lambda_{1,2,3}$ of membrane potentials $V_{m}, V_{I}$ and $V_{F}$, respectively.
When $C_{m}$ ($C_{I}$) is changed by 50\%, while leaving $C_{F}$ and $C_{I}$ ($C_{m}$) unchanged, it can be seen that the fibroblast has the largest decay rate that is one order of magnitude larger than the others. The method is applied to determine the decay rate of the other cells.}
\label{fig:time_scales}
\end{figure}

Fig.~\ref{fig:time_scales} shows the decay rates of the solutions, sorted in such a way that $\lambda_1 \ge \lambda_2 \ge \lambda_3$. The solutions are determined numerically as a function of the coupling conductance, $G_{p}$. The values of the conductances are $G^{\rm int}_{m} = 0.33$nS, $G^{\rm int}_{I} =0.5$nS and $G^{\rm int}_{F} =1.0$nS. We determine the values of $\lambda_i$ with $C_{m} = 120$pF, $C_{I} = 80$pF and $C_{F} =6.0$pF (shown as $\diamond$ in Fig.~\ref{fig:time_scales}), which corresponds to the values used in the numerical study. We have also determined the values of $\lambda_i$, first reducing $C_{m}$ alone by a factor of $2$ (shown as $\square$ in Fig.~\ref{fig:time_scales}), and then reducing $C_{I}$ alone by a factor of $2$ (shown as $\lhd$).
The values of the three decay rates are separated by an order of magnitude from each other. The highest decay rate, $\lambda_1$, depends neither on the capacitance of the myocyte nor on the capacitance of the ICLCs, and can therefore be associated with the dynamics of the fibroblasts. We also find that that diminishing the capacitance $C_{I}$ ($C_{m}$) by a factor $2$ leads to an increase of the intermediate decay rate $\lambda_2$ ($\lambda_3$) by a factor $\sim 2$, which therefore demonstrates that $\lambda_2$ is associated with the relaxation of ICLCs, and $\lambda_3$ with the relaxation of the myocytes.

\begin{figure}[h]
\centering
\subfigure[]{\includegraphics[width=0.33\textwidth]{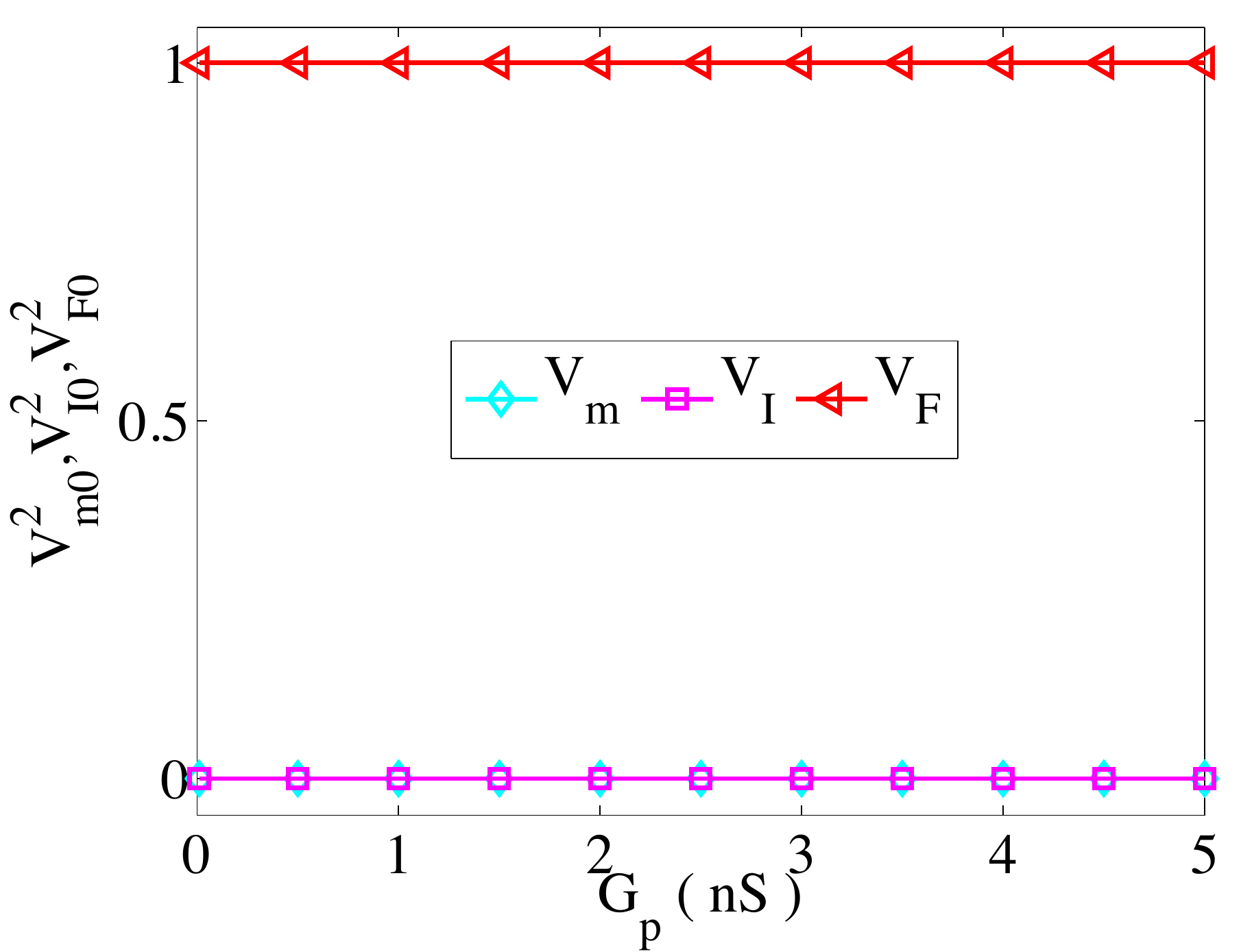}}
\subfigure[]{\includegraphics[width=0.33\textwidth]{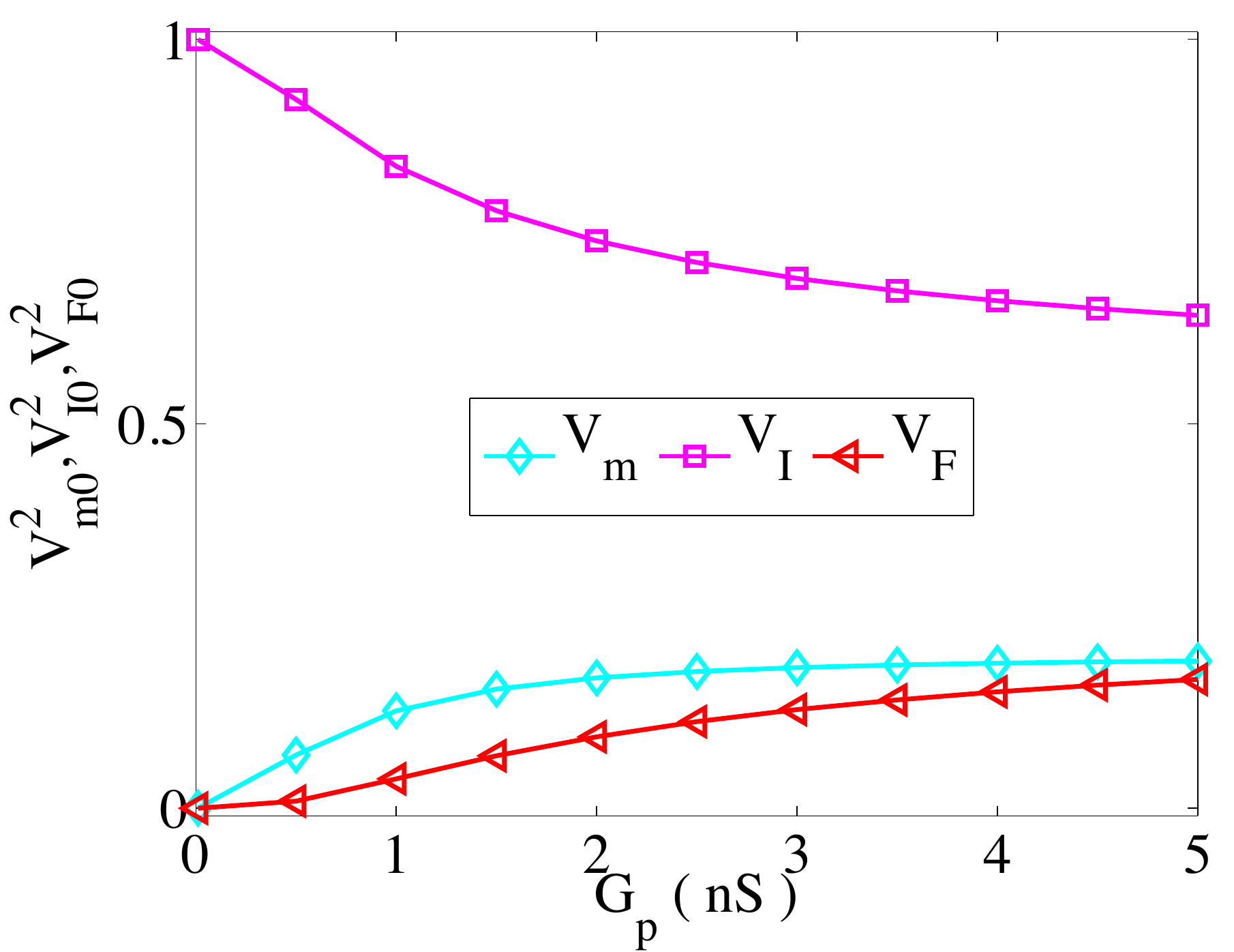}}
\subfigure[]{\includegraphics[width=0.33\textwidth]{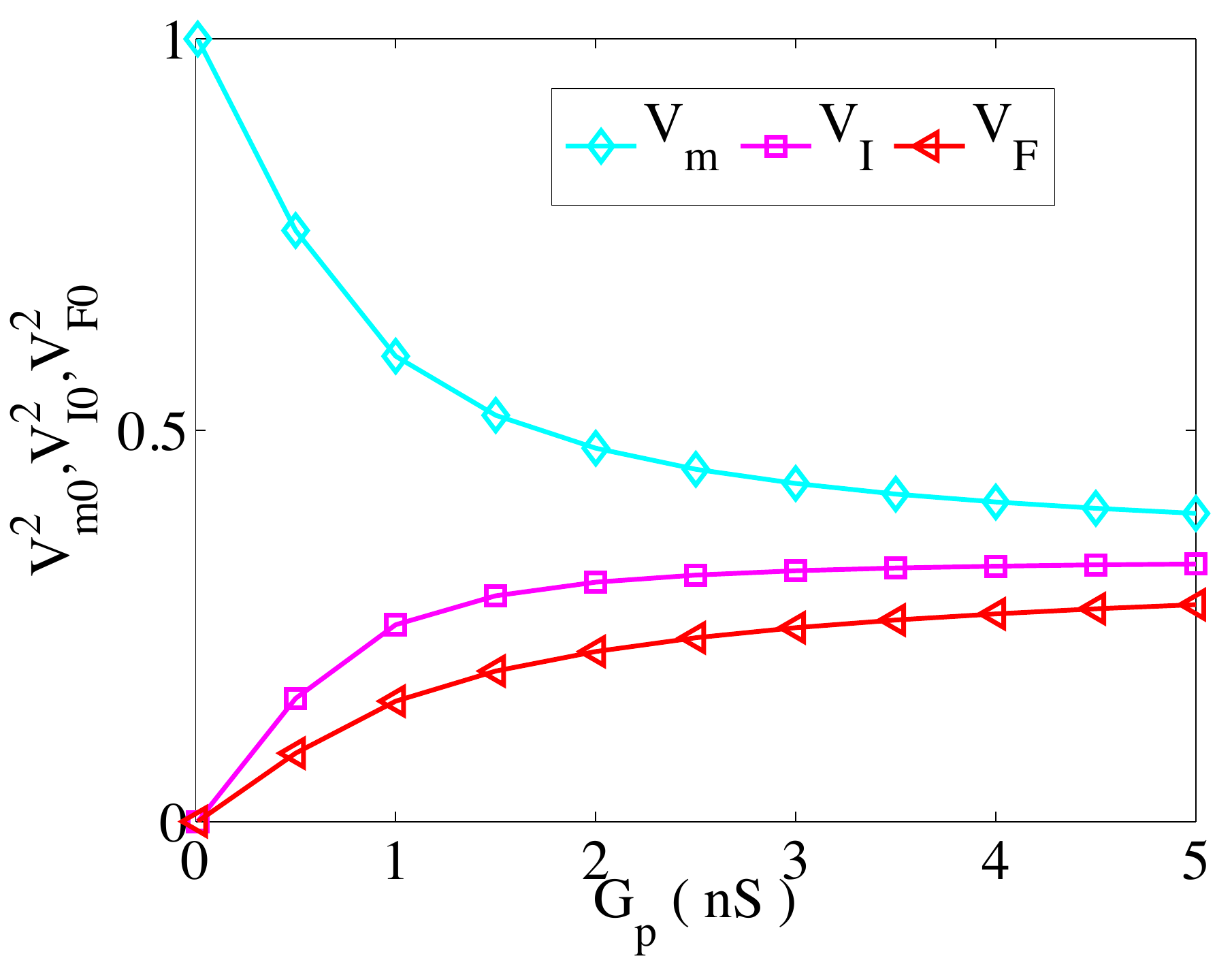}}    
\caption{Components of eigenvectors associated with (a) $\lambda_1$, (b) $\lambda_2$ and (c) $\lambda_3$.}
\label{fig:eigen_vector}
\end{figure}

The alignment of the eigenvectors of the relaxation matrix $M$,
defined by (\ref{eq:def_mat_M}), with the components associated with
the membrane potentials of the fibroblasts, the ICLCs, and
the myocyte, are shown in Fig.~\ref{fig:eigen_vector}~a, b and c
respectively. As the eigenvector associated with the fastest
decaying rate, $-\lambda_1$, is seen to be aligned with $V_{F}$, it
follows that the adaptation of the fibroblast membrane potential
occurs over a much shorter time scale than the other two potentials.
Although the
directions associated with the intermediate (smallest) time scales are
predominantly aligned with the $V_{I}$ ($V_{m}$) directions, the
coupling between the $V_{m}$ and $V_{I}$ components is much stronger,
as indicated by the imperfect alignment between the eigenvectors of
the relaxation matrix and the directions $V_{m}$ and $V_{I}$.

Next, taking into account the fact that passive cells relax quickly, we demonstrate that the dynamics of populations of fibroblasts and ICLCs can be reduced to the dynamics of a population of a single hybrid passive cell type. To this end, equation (\ref{eq:myocyte}) can be rewritten as:
\begin{equation} \label{equ:coupling}
\begin{split}
C_{m}\frac{dV_{m}}{dt}= &-I_{\rm ion} - n_{F}\,G_{p}\,(V_{m}-V_{F}) - n_{I}\,G_{p}\,(V_{m}-V_{I})\\
                      = &-I_{\rm ion} - n_{p}\,G_{p}\,(V_{m}-V_{p})\,,
\end{split}
\end{equation}
where the total number of passive cells coupled to the myocyte $n_{p}=n_{F}+n_{I}$, and where
\begin{equation}
V_{p} = \frac{n_{F}}{n_{p}}V_{F}+\frac{n_{I}}{n_{p}}V_{I}\,,
\label{eq:hybrid}
\end{equation}
is the membrane potential of an effective or ``hybrid" passive cell.

As the fibroblast membrane potential $V_{F}$ evolves on a time scale faster than that of the others (by two orders of magnitude, see Figure \ref{fig:time_scales}), we adiabatically eliminate this variable by using $\frac{dV_{F}}{dt}=0$ in (\ref{eq:fibro}). Replacing $V_{I}$ by $V_{p}$ using (\ref{eq:hybrid}), and differentiating with respect to time, Eq.~(\ref{eq:ICLC}) can be rewritten as
\begin{equation}\label{eq:coup_hyb}
C_{p}\frac{\dd V_{p}}{\dd t} = G_{p}^{\rm int}(V_{p}^{r}-V_{p}) + G_{p}(V_{m}-V_{p})\,,
\end{equation}
which is expressed in terms of the effective parameters
\begin{eqnarray}{\label{equ:hybrid_passive}}
C_{p} &=& \frac{C_{I}}{1+\mu}\,,  \nonumber \\
G^{\rm int}_{p} &=& \frac{G^{\rm int}_{I} - \mu G_{p}}{1+\mu} \,, \nonumber \\
V_{p}^{r} &=& \frac{G^{\rm int}_{I} V_{p}^{rs}-\mu G_{p}V_{F}^{r}}{G^{\rm int}_{I} - \mu G_{p}}\,,
\end{eqnarray}
with $\mu = \frac{n_{F}}{n_{p}}\frac{G^{\rm int}_{I}-G^{\rm int}_{F}}{G_{p}+G^{\rm int}_{F}}$ and $V_{p}^{rs} =\frac{1}{n_{p}}(n_{I} V_{I}^{r} + n_{F} V_{F}^{r})$. In the symmetrical situation where $G^{\rm int}_{I} =G^{\rm int}_{F}$, which is compatible with experimental measurements \cite{Kohl1994,Duquette2005}, $\mu=0$. This simplifies the expressions above, in particular the effective resting potential of the hybrid cell, $V_{p}^{r} = V_{p}^{rs} = \frac{1}{n_{p}}(n_{F}V_{F}^{r}+n_{I}V_{I}^{r})$.


\subsection{Validation}
\label{sub:S4.2}

\begin{figure}[h]
\centering
\includegraphics[width=0.5\textwidth]{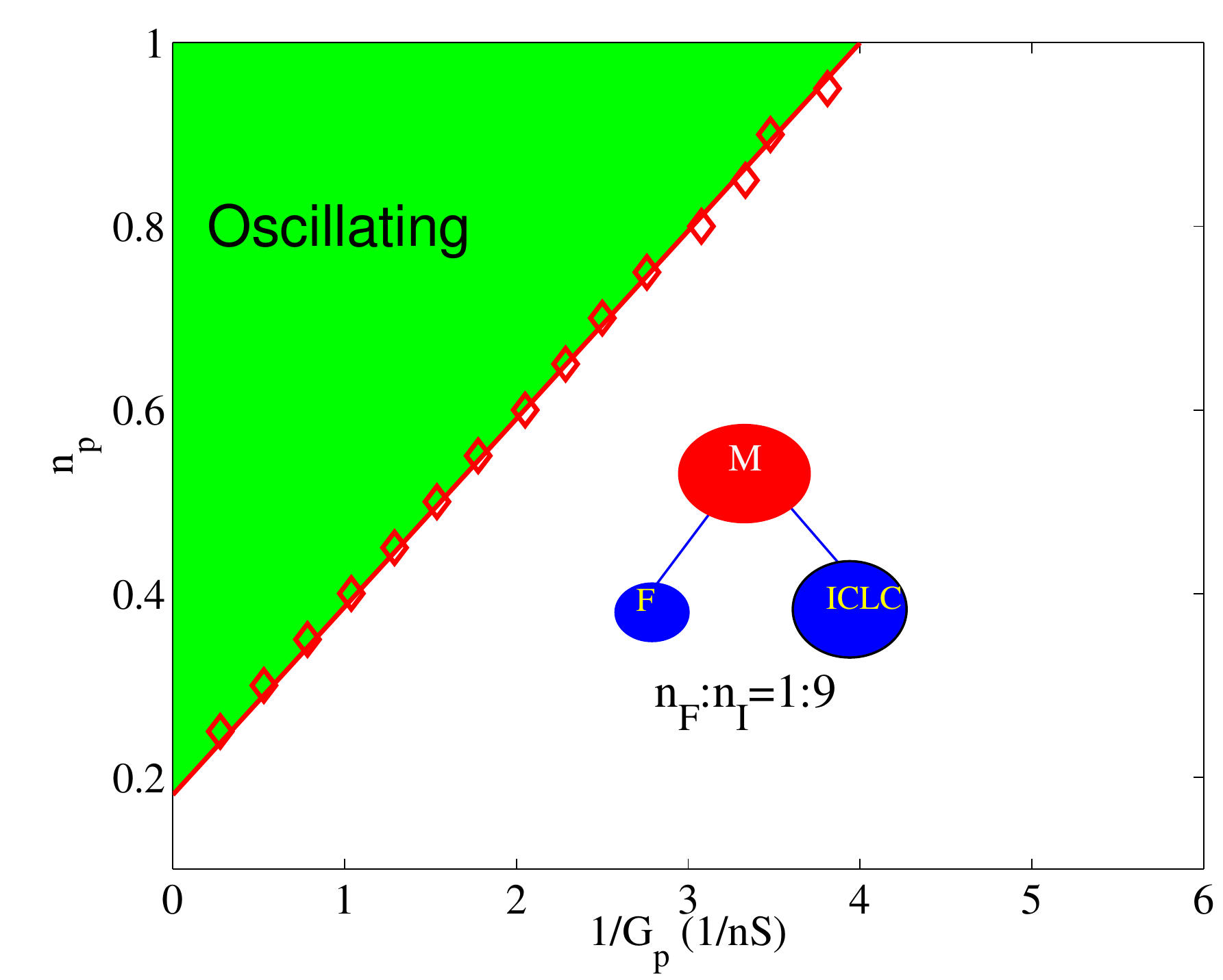}
\caption{
The $(n_{p},G_{p})$ parameter space for the case of a single myocyte coupled to $n_{I}$ ICLCs and $n_{F}$ fibroblasts, indicating the region where oscillations are observed. The ratio of fibroblasts to ICLCs, $nF$:$nI$ is $1$:$9$. For comparison with the results of coupling a myocyte with an effective passive cell see Figure~4 in the main text.
}
\label{fig:fibroblast_influence}
\end{figure}

In order to verify our analysis, we consider the combined effect of ICLCs and fibroblasts on the myocyte dynamics (Eq. \ref{equ:couple}) and compare this situation with the case of a myocyte coupled to $n_{p}$ effective passive cells, described by equation (\ref{eq:hybrid}, \ref{eq:coup_hyb}). We take $n_{p}=n_{F}+n_{I}$ and $n_{F}:n_{I}=1/9$  to describe the small fraction of fibroblasts in the uterine wall \cite{Duquette2005}.  For the chosen set of values, the hybrid passive cell resting potential is $V_{p}^{r}=\frac{1}{n_{p}}(n_{F}V_{F}^{r}+n_{I}V_{I}^{r})\approx-35$mV. Fig.~\ref{fig:fibroblast_influence} shows the range of values of ($1/G_p, n_p$) predicted by Eq.(\ref{equ:couple}) for which the system of myocytes coupled to both ICLCs and fibroblasts oscillates. This should be compared with the range predicted by the model of an effective cell [Eq.(\ref{eq:hybrid},\ref{eq:coup_hyb})], shown in Fig.~4~(b) in the main text. The predictions from the two sets of equations concerning the lowest value of $n_p$, above which oscillations are possible, agree qualitatively. 
In principle, a more refined calculation that does not require the criterion that $C_F/C_I \ll n_F/n_p$ can provide an improved estimate of the parameters for the effective passive cell. However, this approach will probably only result in a minor change in the estimate, given the many uncertainties in the determination of precise parameter values from the available experimental data. Thus, the use of a reduced model for an effective cell, combining ICLC and passive cells with a short relaxation time and a high resting potential, is justified.
%


\end{document}